\documentclass[preprint,5p,times,twocolumn,authoryear]{elsarticle}

\usepackage{amssymb}

\usepackage[utf8]{inputenc}
\usepackage[linesnumbered,lined,ruled]{algorithm2e}
\usepackage{booktabs}
\usepackage{tabularx}
\newcolumntype{Y}{>{\centering\arraybackslash}X}
\newcolumntype{Z}{>{\raggedleft\arraybackslash}X}
\usepackage{multirow}
\usepackage{listings}
\usepackage{fancyvrb}
\usepackage{enumitem}
\usepackage{caption}
\usepackage{subcaption}
\usepackage{hyperref}
\usepackage{stfloats}
\usepackage{xcolor}

\usepackage{todonotes}

\usepackage{pifont}%
\newcommand{\cmark}{\ding{51}}%
\newcommand{\xmark}{\ding{55}}%

\newcommand{\ToolDSE}{DSE}
\newcommand{\ToolSE}{PRV}
\newcommand{\ToolPRV}{PRV}

\newcommand{\ToolARDiff}{ARD\textsc{iff}}

\newcommand{\ToolPASDA}{\textsc{PASDA}}

\newcommand{\scReachable}{\textsc{Reachable}}
\newcommand{\scUnreachable}{\textsc{Unreachable}}
\newcommand{\scMaybeReachable}{\textsc{Maybe\_Reachable}}

\newcommand{\scEq}{\textsc{Eq}}
\newcommand{\scNeq}{\textsc{Neq}}

\newcommand{\scMaybeEq}{\textsc{Maybe\_Eq}}
\newcommand{\scMaybeNeq}{\textsc{Maybe\_Neq}}
\newcommand{\scUnknown}{\textsc{Un\-known}}

\newcommand{\scError}{\textsc{Error}}
\newcommand{\scDepthLimited}{\textsc{Depth\_Lim\-it\-ed}}
\newcommand{\scTimeout}{\textsc{Timeout}}

\newcommand{\scSat}{\textsc{Sat}}
\newcommand{\scUnsat}{\textsc{Unsat}}

\newcommand{\scUif}{\textsc{Uif}}
\newcommand{\scPc}{\textsc{Pc}}

\newcommand{\eqOneNoRef}{\texttt{eq\_v1}}
\newcommand{\eqTwoNoRef}{\texttt{eq\_v2}}
\newcommand{\neqOneNoRef}{\texttt{neq\_v1}}
\newcommand{\neqTwoNoRef}{\texttt{neq\_v2}}

\newcommand{\eqOne}{\hyperref[lst:eq-programs]{\eqOneNoRef}}
\newcommand{\eqTwo}{\hyperref[lst:eq-programs]{\eqTwoNoRef}}
\newcommand{\neqOne}{\hyperref[lst:neq-programs]{\neqOneNoRef}}
\newcommand{\neqTwo}{\hyperref[lst:neq-programs]{\neqTwoNoRef}}

\newcommand{\checkEquivalence}{\hyperref[lst:product-program]{\texttt{check\-Equiv\-a\-lence}}}

\journal{Journal of Systems and Software}

\begin{document}

\begin{frontmatter}

\title{PASDA: A Partition-based Semantic Differencing Approach\\with Best Effort Classification of Undecided Cases}

\author[inst1]{Johann Glock\corref{cor1}}
\ead{johann.glock@aau.at}
\author[inst2]{Josef Pichler}
\author[inst1]{Martin Pinzger}

\cortext[cor1]{Corresponding author.}

\affiliation[inst1]{organization={Department of Informatics Systems, University of Klagenfurt},%
            addressline={Universitätsstraße~65-67}, 
            city={Klagenfurt},
            postcode={9020}, 
            country={Austria}}

\affiliation[inst2]{organization={School of Informatics, Communications and Media, University of Applied Sciences Upper Austria},%
            addressline={Softwarepark 11}, 
            city={Hagenberg im Mühlkreis},
            postcode={4232}, 
            country={Austria}}

\begin{abstract}
    Equivalence checking is used to verify whether two programs produce equivalent outputs when given equivalent inputs. Research in this field mainly focused on improving equivalence checking accuracy and runtime performance. However, for program pairs that cannot be proven to be either equivalent or non-equivalent, existing approaches only report a classification result of \emph{unknown}, which provides no information regarding the programs' non-/equivalence.

    \vspace{.75em}

    In this paper, we introduce \ToolPASDA{}, our partition-based semantic differencing approach with best effort classification of undecided cases. While \ToolPASDA{} aims to formally prove non-/equivalence of analyzed program pairs using a variant of differential symbolic execution, its main novelty lies in its handling of cases for which no formal non-/equivalence proof can be found. For such cases, \ToolPASDA{} provides a best effort equivalence classification based on a set of classification heuristics.

    \vspace{.75em}

    We evaluated \ToolPASDA{} with an existing benchmark consisting of 141 non-/equivalent program pairs. \ToolPASDA{} correctly classified 61--74\% of these cases at timeouts from 10 seconds to 3600 seconds. Thus, \ToolPASDA{} achieved equivalence checking accuracies that are 3--7\% higher than the best results achieved by three existing tools. Furthermore, \ToolPASDA{}'s best effort classifications were correct for 70--75\% of equivalent and 55--85\% of non-equivalent cases across the different timeouts.

\end{abstract}


\end{frontmatter}

\newpage{}
\newpage{}

\section{Introduction}
\label{sec:introduction}

In the context of software programs, the goal of functional equivalence checking is to verify whether two programs produce equivalent outputs when given equivalent inputs \citep{godlin2009regression}. This information can be used, for example, to verify compiler optimizations \citep{dahiya2017black}, to provide assurance that refactorings do not introduce unintended functional changes \citep{person2008differential}, to check whether changes in libraries affect their clients \citep{mora2018client}, or even for security related analyses such as distinguishing benign integer overflows from harmful ones \citep{sun2016inteq} and classifying which malware family a given program belongs to \citep{mercaldo2021equivalence}.

\vspace{.5em}

Classification accuracy and runtime performance of equivalence checking approaches have seen continuous improvements throughout the years \citep{person2011directed, backes2013regression, felsing2014automating, jakobs2022peqtest}, which has enabled these approaches to provide non-/equivalence proofs for more complex programs in shorter amounts of time.
However, for cases that cannot be proven to be either equivalent or non-equivalent, existing approaches such as UC-KLEE \citep{ramos2011practical}, PEQ\textsc{check} \citep{jakobs2021peqcheck}, and \ToolARDiff{} \citep{badihi2020ardiff} only output a classification of \scUnknown{}, which provides no indication about the potential non-/equivalence of the two analyzed programs. For example, \ToolARDiff{} classifies the two equivalent programs shown in Listing~\ref{lst:eq-programs} as well as the two non-equivalent programs shown in Listing~\ref{lst:neq-programs} as \scUnknown{}, but does not provide any further information beyond this. 

In this paper, we introduce \ToolPASDA{}, our \underline{pa}rtition-based \underline{s}emantic \underline{d}ifferencing \underline{a}pproach with best effort classification of undecided cases. While \ToolPASDA{} aims to formally prove non-/equivalence of analyzed program pairs, its main novelty lies in its handling of cases for which no such formal proof can be found. More specifically, \ToolPASDA{} uses a set of classification heuristics to provide a best effort equivalence classification for any cases which cannot be proven to be either equivalent or non-equivalent. For example, \ToolPASDA{} classifies the two equivalent programs in Listing~\ref{lst:eq-programs} as \scMaybeEq{} rather than \scUnknown{}. For the two non-equivalent programs in Listing~\ref{lst:neq-programs}, \ToolPASDA{} can even provide a non-equivalence proof, thus classifying them as \scNeq{} rather than \scUnknown{}. Overall, \ToolPASDA{} distinguishes eight program-level equivalence classifications: equivalent (\scEq{}), non-equivalent (\scNeq{}), maybe equivalent (\scMaybeEq{}), maybe non-equivalent (\scMaybeNeq{}), unknown (\scUnknown{}), depth-limited (\scDepthLimited{}), timeout (\scTimeout{}), and error (\scError{}). 

To further improve the utility of its outputs, \ToolPASDA{} provides additional supporting information along with its program-level equivalence classification results. Specifically, for each input partition that \ToolPASDA{} identifies during a symbolic execution~\citep{king1976symbolic} of the two target programs, it aims to collect (a) concrete and symbolic input and output values, (b) a partition-level equivalence classification, and (c) the lines of code that are covered by the corresponding execution path. Such information has been found to benefit developers' understanding of program analysis results \citep{latoza2010hard,parnin2011automated,winter2022developers}. For this reason, similar information is provided by some existing equivalence checking approaches such as \ToolDSE{}~\citep{person2008differential}, SymDiff~\citep{lahiri2012symdiff}, and \ToolPRV{}~\citep{bohme2013partition}.

{
\definecolor{light-gray}{gray}{0.80}

\lstset{numbers=left,
captionpos=b,
xrightmargin=1.0ex,
xleftmargin=5.0ex,
}

\begin{lstlisting}[language=Java, float=t, escapechar=!, caption=Example of two equivalent programs., label={lst:eq-programs}]
double eq_v1(int x, double y) {
  for (int i = 0; i < !\colorbox{light-gray}{1}!; i++) { x !\colorbox{light-gray}{+= 0}!; }
  return x + y !\colorbox{light-gray}{+ 1}!;
}

double eq_v2(int x, double y) {
  for (int i = 0; i < !\colorbox{light-gray}{x}!; i++) { x !\colorbox{light-gray}{*= 1}!; }
  return !\colorbox{light-gray}{1 +}! x + y;
}
\end{lstlisting}

\begin{lstlisting}[language=Java, float=t, escapechar=!, caption=Example of two non-equivalent programs., label={lst:neq-programs}]
double neq_v1(double x) {
  if (x <= 0) { return !\colorbox{light-gray}{1}!; }
  if (x >  9) { return Math.tan(!\colorbox{light-gray}{1}! * x); }
  return Math.tan(!\colorbox{light-gray}{1}! * x) < 0 ? -1 : 0;
}

double neq_v2(double x) {
  if (x <= 0) { return !\colorbox{light-gray}{2}!; }
  if (x >  9) { return Math.tan(!\colorbox{light-gray}{2}! * x); }
  return Math.tan(!\colorbox{light-gray}{2}! * x) < 0 ? -1 : 0;
}
\end{lstlisting}
}

Our implementation of \ToolPASDA{} takes the source code of two Java programs as its input and first constructs a product program \citep{barthe2011relational,beckert2018trends} that combines the behavior of these two target programs. \ToolPASDA{} then uses the symbolic execution engine Symbolic Pathfinder \citep{pasareanu2010symbolic} to symbolically execute the product program. For each partition identified during the symbolic execution, \ToolPASDA{} collects the corresponding path condition and symbolic output values of the two target programs. This information is sent to the theorem prover Z3 \citep{demoura2008z3} to produce an equivalence classification for each partition. Once all partitions have been analyzed, \ToolPASDA{} aggregates the \emph{partition-level} results to produce a \emph{program-level} result. If the program-level result is a proof of non-/equivalence, \ToolPASDA{} reports this as the final result of its analysis. Otherwise, \ToolPASDA{} employs iterative abstraction and refinement of unchanged parts of the source code \citep{badihi2020ardiff} to conduct further analysis iterations until a non-/equivalence proof is found or a given timeout is reached.

We evaluated the equivalence checking accuracy and runtime performance of \ToolPASDA{} with an existing benchmark~\citep{badihi2020ardiff} consisting of 73 equivalent and 68 non-equivalent program pairs and compared the results to the three existing tools \ToolARDiff{}~\citep{badihi2020ardiff}, \ToolDSE{}~\citep{person2008differential}, and \ToolPRV{}~\citep{bohme2013partition}. \ToolPASDA{} correctly classified 61–74\% of analyzed cases at six different timeout settings ranging from 10 seconds (s) to 3600~s. In our evaluation, \ToolPASDA{}'s classification accuracy is, therefore, 3--7\% higher than those that we observed for the three existing tools, which correctly classified at most 58\% of cases at the 10~s timeout setting and at most 67\% of cases at the 3600~s timeout setting. Furthermore, \ToolPASDA{}'s best effort classifications for undecided cases were correct for 70-75\% of equivalent cases and 55-85\% of non-equivalent cases across the six analyzed timeout settings.

The contributions that we make in this paper are as follows:

\begin{enumerate}
    \item \ToolPASDA{}, an equivalence checking approach for software programs that provides best effort classifications for cases that cannot be proven to be non-/equivalent,
    \item an evaluation of \ToolPASDA{}'s equivalence checking accuracy and runtime performance by comparing it to three state-of-the-art equivalence checking tools using an established equivalence checking benchmark,
    \item an evaluation of \ToolPASDA{}'s best effort classification results,
    \item a publicly available implementation of \ToolPASDA{} that supports equivalence checking of Java programs.
\end{enumerate}

Throughout the remainder of this paper, we first present background information on the use of symbolic execution for equivalence checking in Section~\ref{sec:background}. In Section~\ref{sec:our-approach}, we describe \ToolPASDA{} and evaluate it in Section~\ref{sec:evaluation}. Section~\ref{sec:discussion} discusses the benefits of our approach as well as potential threats to the validity of our results. We present related work in Section~\ref{sec:related-work} and draw the conclusions in Section~\ref{sec:conclusions}.

\section{Background}
\label{sec:background}

This section introduces relevant terminology that we use throughout the rest of this paper to describe our approach. Furthermore, it provides descriptions of the three equivalence checking approaches \ToolDSE{} \citep{person2008differential}, \ToolARDiff{} \citep{badihi2020ardiff}, and \ToolPRV{} \citep{bohme2013partition} which originally introduced key ideas (differential symbolic execution, iterative abstraction and refinement, and partition-based regression verification) that we reuse in our approach.

\subsection{Symbolic Execution}
\label{sec:symbolic-execution}

Symbolic execution is a technique for systematically exploring all possible execution paths of a given program \citep{baldoni2018survey, cadar2013symbolic}. The main ideas employed by symbolic execution are to (i) use symbolic rather than concrete values to represent the program state and to (ii) follow all reachable execution paths whenever a branching point is encountered during the execution \citep{king1976symbolic}. For each execution path, symbolic execution collects a \textit{path condition} that contains the constraints that have to be satisfied by the input variables to reach the path, and the outputs produced by following the path, which are represented as a function of the input variables \citep{pasareanu2009survey}. The input values that satisfy a given path condition are commonly referred to as an \textit{(input) partition}.

Once all reachable paths have been explored, symbolic execution outputs a \textit{program summary} which consists of a disjunction of one or more \textit{partition-effects pairs} \citep{person2008differential}. Each partition-effects pair consists of a conjunction of the path condition and corresponding outputs of a single execution path. Unreachable paths, i.e., paths that have unsatisfiable path conditions, are skipped during the symbolic execution and not included in the program summaries. For example, the program summary that is produced by a symbolic execution of the program \neqTwo{} in Listing~\ref{lst:neq-programs} is the following:

\begin{lstlisting}[aboveskip=1em,belowskip=1em,mathescape=true]
    x<=0                           $\land$ RET=2
$\lor$ !(x<=0) $\land$     x>9                  $\land$ RET=tan(2*x)
$\lor$ !(x<=0) $\land$ !(x>9) $\land$      tan(2*x)<0  $\land$ RET=-1
$\lor$ !(x<=0) $\land$ !(x>9) $\land$ !(tan(2*x)<0) $\land$ RET=0 
\end{lstlisting}

Limitations of symbolic execution include its high computational cost, its inability to produce complete summaries in the presence of unbounded loops and recursion, and its inability to handle complex expressions (e.g., non-linear arithmetic) which are intractable for modern decision procedures \citep{person2008differential, cadar2011symbolic, cadar2013symbolic}.

\subsection{Differential Symbolic Execution}
\label{sec:dse}

Differential symbolic execution (DSE) uses symbolic execution to characterize differences in the input-output behavior of two programs \citep{person2008differential}. More specifically, DSE first collects summaries of the two target programs via symbolic execution. It then compares these summaries to identify inputs for which the two programs produce different outputs. If no differences in the input-output behavior are identified, the two programs are equivalent and DSE does not produce any further output. However, if the two programs are non-equivalent, DSE outputs a \textit{behavioral delta} that consists of the non-equivalent input partitions and corresponding outputs that are produced for these inputs by the two target programs. 

To avoid some of the limitations of symbolic execution, DSE replaces parts of the source code that are unchanged across the two target programs with calls to uninterpreted functions (UIFs) before running the symbolic execution.
For example, assuming that line 2 of \eqOne{} is syntactically equivalent to line 7 of \eqTwo{} in Listing~\ref{lst:eq-programs}, both lines would be replaced with a UIF call $x = UIF(x)$.
While this abstraction would lead to an overapproximation of actual program behavior, it would enable DSE to prove the two programs to be equivalent without having to bound the number of loop iterations that are analyzed. 
Similar benefits can be achieved when abstracting program constructs such as recursion and non-linear arithmetic that are also difficult for symbolic execution to handle.

One of the main limitations of DSE is that it needs to \emph{fully} analyze both target programs before an equivalence classification can be produced. This makes it unsuitable for use cases where its execution time exceeds given time constraints. Furthermore, DSE exhibits lower equivalence checking accuracy than more recent tools such as \ToolARDiff{} \citep{badihi2020ardiff}. Finally, even though \ToolDSE{} produces behavioral deltas for \scNeq{} partitions, it does not provide input-output descriptions for partitions classified as \scEq{} or \scUnknown{}. For example, \ToolDSE{} classifies the programs in Listing~\ref{lst:eq-programs} as \scUnknown{} due to depth-limiting, and Listing~\ref{lst:neq-programs} as \scUnknown{} due to the presence of the \texttt{tan()} function (which generally is not fully modeled by modern decision procedures), but does not provide any information beyond this.

\subsection{Iterative Abstraction and Refinement}
\label{sec:ardiff}

The goal of \ToolARDiff{}~\citep{badihi2020ardiff} is to reduce the number of programs that cannot be provably classified as non-/equivalent due to the introduction of UIFs while preserving the benefits that this abstraction provides. \ToolARDiff{} accomplishes this through an iterative process %
that starts with the same abstraction of unchanged source code that \ToolDSE{} uses. If no non-/equivalence proof can be found at this level of abstraction, \ToolARDiff{} refines one of the UIFs, replacing it with the original unchanged source code. The resulting partially refined program is again checked for equivalence, and further refinement iterations are conducted until non-/equivalence can be proven, no further refinement is possible, or a timeout is reached.

To choose which UIF should be refined in each iteration, \ToolARDiff{} employs three heuristics. Heuristic 1 (H1) marks those UIFs as refinement candidates for which an assignment exists that enables a non-/equivalence proof irrespective of the assignments of all other UIFs. Heuristic 2 (H2) marks those UIFs as refinement candidates that occur a different number of times in the two programs. Heuristic 3 (H3) then ranks the refinement candidates according to how deeply they are nested in loops and how many non-linear arithmetic operations each candidate replaces. If no refinement candidates are identified by H1 and H2, all UIFs are included in H3's ranking. The UIF with the lowest rank is refined. We refer the reader to ~\citet{badihi2020ardiff} for a more detailed description of the three heuristics.

While \ToolARDiff{} achieves better equivalence checking accuracy than \ToolDSE{}, this comes at the cost of longer runtimes. Furthermore, \ToolARDiff{} inherits \ToolDSE{}'s requirement that the two target programs have to be \emph{fully} analyzed before an equivalence classification can be provided. Finally, \ToolARDiff{} does \textit{not} preserve \ToolDSE{}'s behavioral deltas as a description of non-equivalent program behaviors --- it only produces program-level equivalence classifications. For example, \ToolARDiff{} reports the two cases shown in Listing~\ref{lst:eq-programs} and Listing~\ref{lst:neq-programs} as \scUnknown{} for the same reasons as \ToolDSE{}, but does not provide any other information beyond these equivalence classifications. %

\subsection{Partition-based Regression Verification}
\label{sec:prv}

\ToolDSE{} and \ToolARDiff{} both need to fully analyze the two target programs before a non-/equivalence proof can be provided.
Partition-based regression verification (\ToolPRV{}), on the other hand, can provide non-equivalence proofs for many cases after only a partial analysis of the target programs \citep{bohme2013partition}.
\ToolPRV{} accomplishes this by checking non-/equivalence on the partition level instead of the program level.
With this approach, two programs are proven to be equivalent if \emph{all} partitions can be proven to be so, which still requires a full analysis.
However, two programs are proven to be non-equivalent as soon as the \emph{first} input partition for which the two target programs produce different outputs is identified.

\ToolPRV{}'s analysis starts by identifying a random assignment of input values from the common input space of the two target programs.
These inputs are used to execute both programs, which provides the program outputs and the corresponding path conditions.
Outputs are checked for non-/equivalence, thus marking the partition as either \scEq{} or \scNeq{} if a non-/equivalence proof is found, or \scUnknown{} otherwise.
Exploration then continues with a new set of input values from the unexplored part of the common input space.
The process repeats until the full input space has been explored or some timeout is reached.

There are two main benefits that \ToolPRV{}'s partition-based approach provides:
(i)~non-equivalence of two programs can be proven without a full analysis of all program paths, and
(ii)~even if the analysis is interrupted ahead of time, non-/equivalence proofs for all explored partitions are preserved.
However, \ToolPRV{} does not employ UIFs to abstract unchanged code parts and, therefore, cannot benefit from the advantages that such an abstraction entails (see Section~\ref{sec:dse}).
Furthermore, \ToolPRV{} still distinguishes only three equivalence classifications (\scEq{}, \scNeq{}, and \scUnknown{}) at the partition level. For example, \ToolPRV{} classifies all partitions in Listing~\ref{lst:neq-programs} that return via line 4 in \neqOne{} and line 10 in \neqTwo{} as \scUnknown{} because it cannot generate inputs that satisfy the corresponding path conditions (due to the presence of the \texttt{tan()} function).

\section{The \ToolPASDA{} Approach}
\label{sec:our-approach}

Our approach, which we refer to as \ToolPASDA{}, combines ideas from
\ToolDSE{} \citep{person2008differential} (abstraction of unchanged code parts), 
\ToolARDiff{} \citep{badihi2020ardiff} (iterative abstraction and refinement), and 
\ToolPRV{} \citep{bohme2013partition} (partition-based analysis)
and further extends them with a best effort classification of programs and partitions that cannot be formally proven to be either equivalent or non-equivalent. 

\begin{figure}[tbp]
    \centering
    \includegraphics[keepaspectratio, width=\columnwidth]{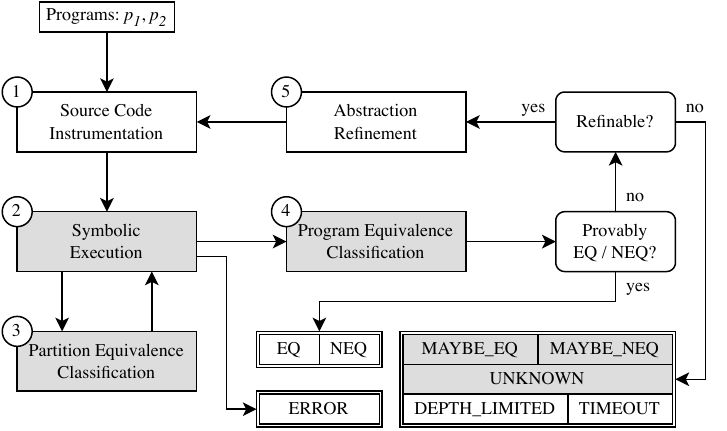}
    \caption{Overview of \ToolPASDA{}'s semantic differencing process. Elements with a white background are identical to the state-of-the-art, whereas elements with a gray background are modified or newly introduced by \ToolPASDA{}.}
    \label{fig:approach-overview}
\end{figure}

An overview of the full approach is shown in Figure~\ref{fig:approach-overview}. \ToolPASDA{} starts with a source code instrumentation step that creates a product program which combines the two target programs $p_1$ and $p_2$. Starting at \ToolPASDA{}'s second iteration, this step also replaces unchanged parts of the source code of the two target programs with UIFs. Once the instrumentation is complete, \ToolPASDA{} symbolically executes the product program to collect information for equivalence checking. If non-/equivalence cannot be proven based on the collected information, the three heuristics proposed by \citet{badihi2020ardiff} are used to choose a UIF to replace with the original unchanged code, thereby creating two new programs that are again checked for non-/equivalence. This iterative refinement process continues until one of the following conditions holds:
(i)~equivalence or non-equivalence can be proven, 
(ii)~no further refinement is possible, or 
(iii)~the runtime exceeds the configured timeout.

The main difference that sets \ToolPASDA{} apart from existing approaches is how it handles limitations of the used decision procedures.
For example, if a program path cannot be proven to be reachable or unreachable, or if the outputs of two programs cannot be proven to be equivalent or non-equivalent, existing tools simply classify the corresponding programs or partitions as \scUnknown{}.
\ToolPASDA{}, on the other hand, takes a more differentiated view.
Rather than classifying all such cases as \scUnknown{}, it distinguishes between \scMaybeEq{}, \scMaybeNeq{}, and \scUnknown{} results based on a set of classification heuristics (see \hyperref[sec:partition-equivalence-classification]{Step~3} and \hyperref[sec:program-equivalence-classification]{Step~4}).
Furthermore, \ToolPASDA{} provides not only program- and partition-level equivalence classifications, but also aims to provide execution traces as well as concrete and symbolic inputs and outputs for identified partitions (see \hyperref[sec:symbolic-execution-step]{Step~2} and Section~\ref{sec:program-output}) since such information has been found to benefit developers' understanding of program analysis results \citep{latoza2010hard,parnin2011automated,winter2022developers}.

In the following, we provide more detailed descriptions of \ToolPASDA{}'s five processing steps in Sections~\ref{sec:source-code-instrumentation}--\ref{sec:abstraction-refinement} as well as its outputs in Section~\ref{sec:program-output}.

\subsection{Step 1: Source Code Instrumentation}
\label{sec:source-code-instrumentation}

At the start of every iteration, \ToolPASDA{} constructs a product program \citep{barthe2011relational,beckert2018trends} that enables it to check the equivalence of the two target programs in a single symbolic execution run \citep{lahiri2012symdiff, ramos2011practical}. Furthermore, starting at its second iteration, \ToolPASDA{} replaces parts of the source code that are unchanged across the two target programs with UIFs \citep{person2008differential}.

\paragraph{Construction of the Product Program}
Listing~\ref{lst:product-program} shows the template that is used by \ToolPASDA{} to construct the product program. The product program executes the two given target programs (lines 15 and 20) and checks whether their effects, i.e., return values or thrown exceptions, are equivalent (lines 25 and 26). If the effects are \textit{not} equivalent, a \texttt{DifferentOutputsException} is thrown. Otherwise, the result produced by the two programs is returned as the output of the product program (line 28).

\begin{lstlisting}[language=Java, float=t, escapechar=!, caption=The product program template used by \ToolPASDA{}., label={lst:product-program}]
public class ProductProgram {
  ...
  
  public static void main(String[] args) {
    ProductProgram.run(${values});
  }
  
  public static ${type} run(${params}) {
    ${type} result1 = ${defaultValue};
    ${type} result2 = ${defaultValue};
    Throwable error1 = null;
    Throwable error2 = null;
    
    try {
      result1 = ${cls1}.${method1}(${vars});
    } catch (Throwable e) {
      error1 = e;
    }
    try {
      result2 = ${cls2}.${method2}(${vars});
    } catch (Throwable e) {
      error2 = e;
    }
    
    checkEquivalence(error1, error2);
    checkEquivalence(result1, result2);
    
    return result2;
  }

  public static void checkEquivalence(...)
    throws DifferentOutputsException { ... }
}
\end{lstlisting}

\bigskip{}
For the product program template to be applicable to a given pair of target programs, the following assumptions must hold:
(i) the two programs must take the same number of input arguments, and the types of the input arguments as well as the output types of the two programs must match, %
(ii) all effects produced by the two target programs must be observable via return values or raised exceptions,
and (iii) the two target programs must both be deterministic, i.e., both programs must always produce equivalent effects when called with equivalent inputs.

In practice, program pairs that do not satisfy assumption (i) could be detected and classified as non-equivalent through a lightweight preprocessing step that compares the input and output types of the two target programs.
Furthermore, programs that do not satisfy assumption (ii) could be transformed to include side effects (e.g., modifications of global variables, console outputs, etc.) in the outputs of the programs prior to equivalence checking.
We leave such improvements for future work.

\paragraph{Introduction of UIFs}
Starting at its second iteration, \ToolPASDA{} uses instrumented variants of the two target programs as the targets of its analysis.
In these instrumented programs, parts of the source code that are syntactically unchanged across the two original programs are replaced with UIFs.
We reuse \ToolARDiff{}'s implementation~\citep{badihi2020github} of this instrumentation step, which uses GumTree~\citep{falleri2014fine} to identify unchanged parts of the source code. When applied to our examples in Listing~\ref{lst:eq-programs} and Listing~\ref{lst:neq-programs}, none of the statements are replaced, because each line of code contains a syntactic change.

\subsection{Step 2: Symbolic Execution}
\label{sec:symbolic-execution-step}

\ToolPASDA{} uses Symbolic Pathfinder (SPF) \citep{pasareanu2010symbolic} to symbolically execute the product program created during the instrumentation step.
For each identified path that is either provably or maybe reachable (for further details on reachability, see \emph{Partition Reachability Classification} in Section~\ref{sec:partition-equivalence-classification}),
\ToolPASDA{} aims to collect the following data:

\begin{enumerate}[label=($\alph*$)]
    \item one partition-effects pair for each program version,
    \item equivalence information for the two partition-effects pairs,
    \item information about the lines of code covered by the path.
\end{enumerate}

The ($a$) partition-effects pairs and ($c$) coverage information are collected via listeners that hook into the lifecycle events of SPF. For example, coverage information is collected whenever SPF raises an \texttt{in\-struc\-tion\-Ex\-e\-cu\-ted} event. The ($b$) equivalence information is calculated whenever a \checkEquivalence{} call is reached in lines 25 and 26 of the product program shown in Listing~\ref{lst:product-program}. For exceptions, \ToolPASDA{} only checks whether their types are the same to determine non-/equivalence. For regular return values, \ToolPASDA{} performs the partition equivalence classification described in \hyperref[sec:partition-equivalence-classification]{Step~3}. %
Note that the full data for ($a$)--($c$) can only be collected for paths that (i)~are (maybe) reachable, and (ii)~are not depth-limited.
In the following paragraphs, we describe which parts of the data are collected for the remaining paths that do not satisfy these criteria.

\paragraph{Unreachable Paths} 
A path is said to be unreachable if its path condition is provably unsatisfiable. For example, in \eqOne{} in Listing~\ref{lst:eq-programs}, all paths that perform more than a single loop iteration are unreachable because $i < 1$ cannot be satisfied anymore once $i$ has been incremented. Since unreachable paths represent program flow that cannot occur in practice, all unreachable paths are skipped during the symbolic execution, and \ToolPASDA{} does not collect any of the data listed in ($a$)--($c$) for them.

\paragraph{Depth-Limited Paths} 
For paths that are not fully analyzed due to the configured depth limit, \ToolPASDA{} only collects partial data for (a)--(c).
Specifically, it collects
($a'$) the path condition of the path, and
($c'$) the covered lines of code up to the point where the depth limit was reached. For example, when \eqTwo{} from Listing~\ref{lst:eq-programs} is symbolically executed with a depth limit of $10$, all inputs $x > 10$ produce depth-limited paths. For these inputs, execution proceeds as normal until the program is about to enter the 11th loop iteration. At this point, symbolic execution of the current path stops with a path condition of $x > 10$, having covered only line~2.
No ($b$) equivalence information can be collected because no return value has been produced at this point. The corresponding partition is classified as \scDepthLimited{} and the symbolic execution then proceeds as usual for the remaining paths that have not been explored yet.

\medbreak{}
\ToolPASDA{} alternates between \hyperref[sec:symbolic-execution-step]{Step~2} and \hyperref[sec:partition-equivalence-classification]{Step~3} until all (maybe) reachable paths have been explored and checked for non-/equivalence. The process then continues with the program equivalence classification in \hyperref[sec:program-equivalence-classification]{Step~4}.

\subsection{Step 3: Partition Equivalence Classification}
\label{sec:partition-equivalence-classification}

As described in the previous section, the values returned by the two target programs are checked for equivalence whenever the symbolic execution reaches the \checkEquivalence{} call in line 26 of the product program shown in Listing~\ref{lst:product-program}. Two factors are combined to produce the overall partition-level equivalence classification: partition reachability and partition output equivalence. In this section, we describe how these factors are calculated and how the overall partition-level equivalence classification is derived from them. %

\paragraph{Partition Reachability Classification}
Reachability can be classified as \scReachable, \scMaybeReachable, or \scUnreachable. Classification is influenced by two factors: (i) the presence of UIFs in the path condition of the partition (\scUif{}~= yes / no), and (ii)~the result of a Z3 query that checks whether the path condition is satisfiable (\scPc-Query = \scSat{} / \scUnsat{} / \scUnknown). An enumeration of the possible combinations of these factors and their corresponding classifications are shown in Table~\ref{tab:partition-reachability}.
For example, paths in \neqTwo{} from Listing~\ref{lst:neq-programs} that return via line 8 or 9 are \scReachable{}. Paths that return via line 10 are \scMaybeReachable{} because Z3 cannot prove whether $Math.tan(2 * x) < 0$ holds for all $x$. Paths in \eqOne{} from Listing~\ref{lst:eq-programs} that perform more than one loop iteration are \scUnreachable{}. As described in the previous section, \scUnreachable{} partitions are skipped by the symbolic execution and, therefore, do not have to be considered during partition equivalence classification.

\paragraph{Partition Output Equivalence Classification} 
Output equivalence can be classified as \scEq, \scNeq, \scMaybeEq, \scMaybeNeq, or \scUnknown. Classification is influenced by three factors: (i)~the presence of UIFs in the partition-effects pairs of the current path (\scUif{}~= yes / no), (ii)~the result of a Z3 query that checks whether the outputs of the two programs are non-equivalent (\scNeq-Query~= \scSat{} / \scUnsat{} / \scUnknown), and (iii) the result of a Z3 query that checks whether the outputs of the two programs are equivalent (\scEq-Query~= \scSat{} / \scUnsat{} / \scUnknown). An overview of the possible combinations of these factors and their corresponding classifications are shown in Table~\ref{tab:partition-output-classification}. For example, the path that returns via line 2 in \neqOne{} and line 8 in \neqTwo{} in Listing~\ref{lst:neq-programs} is classified as \scNeq{}, whereas the path that returns via line 3 in \neqOne{} and line 9 in \neqTwo{} is classified as \scUnknown{} because Z3 cannot prove whether or not $Math.tan(1 * x)$ and $Math.tan(2 * x)$ are equivalent for all $x$.

\paragraph{Overall Partition Equivalence Classification} 
The partition reachability classification and partition output classification are combined to produce the overall partition-level equivalence classification. %
Specifically, a reachability result of \scMaybeReachable{} turns \scNeq{} results into \scMaybeNeq{} results (because in this case, it is not possible to prove that the non-equivalence can be observed in a concrete execution of the two original programs). For all other combinations, the overall classification is the same as the output classification. %
Thus, each partition is assigned to one of six equivalence classes: \scEq{}, \scNeq{}, \scMaybeEq{}, \scMaybeNeq{} or \scUnknown{} if the partition is \textit{not} depth-limited (as described in this section), or \scDepthLimited{} otherwise (as described in the previous section). 
Classification results are returned to \hyperref[sec:symbolic-execution-step]{Step~2} which then continues with the analysis of the remaining 
paths of the product program.

\begin{table}[tbp]
\centering
\caption{Partition reachability classification.}
\label{tab:partition-reachability}
\begin{tabular}{cll} \toprule
\scUif & \scPc-Query & Reachability Classification     \\ \midrule
no   & \scSat  & \scReachable        \\
no   & \scUnsat  & \scUnreachable      \\
no   & \scUnknown  & \scMaybeReachable \\
yes   & \scSat  & \scMaybeReachable \\
yes   & \scUnsat  & \scUnreachable      \\
yes   & \scUnknown  & \scMaybeReachable \\ \bottomrule
\end{tabular}
\end{table}

\begin{table}[tbp]
\centering
\caption{Partition output equivalence classification.}
\label{tab:partition-output-classification}
\begin{tabular}{clll} \toprule
\scUif & \scNeq-Query     & \scEq-Query      & Output Classification \\ \midrule
$\ast$   & \scUnsat   & $\ast$     & \scEq \\
$\ast$   & \scUnknown & \scSat & \scMaybeEq \\
$\ast$   & \scUnknown & \scUnsat & \scNeq \\
$\ast$   & \scUnknown & \scUnknown & \scUnknown \\
no  & \scSat     & $\ast$     & \scNeq                \\
yes & \scSat     & \scSat     & \scMaybeNeq         \\
yes & \scSat     & \scUnsat   & \scNeq                \\
yes & \scSat     & \scUnknown & \scMaybeNeq         \\ \bottomrule
\end{tabular}
\end{table}

\subsection{Step 4: Program Equivalence Classification}
\label{sec:program-equivalence-classification}

Once the symbolic execution concludes, the results of the partition-level equivalence classifications are aggregated to form the program-level equivalence classification of the current iteration. A program-level result of \scEq{} is reported if all partitions could be identified, analyzed, and proven to be \scEq{} within the given timeout. A result of \scMaybeEq{} is reported if at least some partitions were identified and classified as \scEq{} or \scMaybeEq{} but none were classified as \scNeq{} or \scMaybeNeq{}. Otherwise, the program-level result is determined by the partition-level result with the highest priority, where priorities are as follows: \scNeq{} \textgreater{} \scMaybeNeq{} \textgreater{} \scUnknown{} \textgreater{} \scDepthLimited{}. %
For example, Table~\ref{tab:partition-level-data} lists the partition-level data obtained by \ToolPASDA{} for \neqOne{} and \neqTwo{} in Listing~\ref{lst:neq-programs}. Because at least one partition is classified as \scNeq{}, the two programs are classified as \scNeq{}. If an error (e.g., \texttt{OutOfMemoryError}) occurs during \ToolPASDA{}'s execution, a classification of \scError{} is reported. If \ToolPASDA{} cannot analyze even a single partition within the given timeout, a classification of \scTimeout{} is reported.

\subsection{Step 5: Abstraction Refinement}
\label{sec:abstraction-refinement}

Abstraction refinement takes place at the end of an iteration if neither equivalence nor non-equivalence can be proven. Refinement, in this case, refers to the process of selecting one of the UIFs that were introduced during the instrumentation step and marking it to be excluded from abstraction in all following iterations. Consequently, the number of UIFs that remain in the instrumented programs is reduced by one for every iteration that concludes with a program-level equivalence classification that is neither \scEq{} nor \scNeq{}. 
We are reusing \ToolARDiff{}'s implementation~\citep{badihi2020github} of the abstraction refinement step that applies the three heuristics described in Section~\ref{sec:ardiff}. For a more detailed description of the heuristics, we refer the reader to \ToolARDiff{} \citep{badihi2020ardiff}.

\begin{table*}
{
\caption{Partition-level data collected by \ToolPASDA{} when checking equivalence of \neqOne{} and \neqTwo{} shown in Listing~\ref{lst:neq-programs}.}
\label{tab:partition-level-data}
\setlength{\tabcolsep}{5.4pt}
\begin{tabularx}{\textwidth}{Xlllrrlll} \toprule
\# & Path Condition & \multicolumn{2}{l}{Covered Lines} & \multicolumn{2}{r}{Output} & Reachability   & Output & Overall \\ \cmidrule(l{2pt}r{2pt}){3-4} \cmidrule(l{2pt}r{2pt}){5-6}
&               & v1 & v2                           & v1 & v2                    & Classification & Class. & Class.  \\ \midrule
1 & $x < 0$                                                & 2       & 8        & $1$      & $2$       & \scReachable{}      & \scNeq{}     & \scNeq{}      \\
2 & $x = 0$                                                & 2       & 8        & $1$      & $2$       & \scReachable{}      & \scNeq{}     & \scNeq{}      \\
3 & $x > 0 \land x < 9 \land tan(x) < 0 \land tan(2x) < 0$ & 2, 3, 4 & 8, 9, 10 & $-1$     & $-1$      & \textsc{Maybe\_Reach.} & \scEq{}      & \scEq{}       \\
4 & $x > 0 \land x < 9 \land tan(x) < 0 \land tan(2x) > 0$ & 2, 3, 4 & 8, 9, 10 & $-1$     & $0$       & \textsc{Maybe\_Reach.} & \scNeq{}     & \scMaybeNeq{} \\
5 & $x > 0 \land x < 9 \land tan(x) > 0 \land tan(2x) < 0$ & 2, 3, 4 & 8, 9, 10 & $0$      & $-1$      & \textsc{Maybe\_Reach.} & \scNeq{}     & \scMaybeNeq{} \\
6 & $x > 0 \land x < 9 \land tan(x) > 0 \land tan(2x) > 0$ & 2, 3, 4 & 8, 9, 10 & $0$      & $0$       & \textsc{Maybe\_Reach.} & \scEq{}      & \scEq{}       \\
7 & $x > 0 \land x = 9 \land tan(x) < 0 \land tan(2x) < 0$ & 2, 3, 4 & 8, 9, 10 & $0$      & $0$       & \textsc{Maybe\_Reach.} & \scEq{}      & \scEq{}       \\
8 & $x > 0 \land x > 9                                   $ & 2, 3 & 8, 9        & $tan(x)$ & $tan(2x)$ & \scReachable{}      & \scUnknown{} & \scUnknown{}  \\ \bottomrule
\end{tabularx}
}
\end{table*}

\subsection{Program Output}
\label{sec:program-output}

The data that is collected by \ToolPASDA{} is stored in an SQLite database to make it easily available for further 
processing and analysis. For example, Table~\ref{tab:partition-level-data} shows a subset of the partition-level data that \ToolPASDA{} collects when checking equivalence of \neqOne{} and \neqTwo{} from Listing~\ref{lst:neq-programs}. As described in Section~\ref{sec:symbolic-execution-step}, the collected data contains ($a$)~partition-effects pairs for the two program versions (columns \textit{Path Condition} and \textit{Output v1/v2}), ($b$)~equivalence information (columns \textit{Reachability Classification}, \textit{Output Classification}, and \textit{Overall Classification}), and ($c$)~information about the lines of code that are reached when executing the two target programs with corresponding input values (columns \textit{Covered Lines v1/v2}).

For runs that finish within the configured timeout, \ToolPASDA{} always reports the data collected during its last iteration as the result of its overall analysis. For runs that do \textit{not} finish within the timeout, \ToolPASDA{} either reports the data of (i) the last iteration $i_n$ (which \textit{did} time out) or (ii) the second-to-last iteration $i_{n-1}$ (which did \textit{not} time out) based on the following criteria: if $i_n$ is the only iteration or produces a non-equivalence proof, the data of $i_n$ is reported. Otherwise, the data of $i_{n-1}$ is reported. This special logic is applied to avoid situations where little to no data would be reported when a timeout occurs shortly after $i_n$ is started. Note that an \scEq{} result can never be produced for timed-out iterations because equivalence can only be proven if all partitions are analyzed.

\section{Evaluation}
\label{sec:evaluation}

The goals of our evaluation are (i)~to compare the equivalence checking accuracy and runtime performance of \ToolPASDA{} to existing, state-of-the-art equivalence checking approaches, and (ii)~to measure the accuracy of \ToolPASDA{}'s best effort equivalence classification for undecided cases, i.e., program pairs for which no non-/equivalence proof can be found. In line with these goals, we define the following research questions:

\begin{itemize}
    \item \textbf{RQ1}: What is the program-level equivalence classification accuracy of \ToolPASDA{} compared to existing equivalence checking approaches?
    \item \textbf{RQ2}: How accurate are \ToolPASDA{}'s best effort equivalence classifications of undecided cases?
    \item \textbf{RQ3}: How many partitions are classified as (maybe) non-/equivalent by PASDA compared to PRV?
    \item \textbf{RQ4}: What is the runtime performance of \ToolPASDA{} compared to existing equivalence checking approaches?
\end{itemize}

In the following, we first describe the tools included in our evaluation in Section~\ref{sec:evaluated-tools} and the benchmark cases on which the evaluation was performed in Section~\ref{sec:benchmark-programs}. In Sections~\ref{sec:program-classification-results}--\ref{sec:runtime-performance-results}, we describe the evaluation results corresponding to the four research questions RQ1--RQ4.

\subsection{Evaluated Tools / Approaches}
\label{sec:evaluated-tools}

The equivalence checking approaches that we included in our evaluation are \ToolPASDA{}, \ToolARDiff{} \citep{badihi2020ardiff}, \ToolDSE{} \citep{person2008differential}, and \ToolPRV{} \citep{bohme2013partition}. Table~\ref{tab:properties-of-approaches} shows an overview of the main properties along which these approaches can be differentiated. \ToolARDiff{} and \ToolDSE{} are both summary-based tools that introduce uninterpreted functions (UIFs) as an abstraction for unchanged parts of the source code. Furthermore, \ToolARDiff{} iteratively refines these abstractions to mitigate their drawbacks while preserving their benefits at the cost of increased runtimes. \ToolPRV{}, on the other hand, is a partition-based approach that does \emph{not} introduce UIFs into the analyzed programs. Our own approach, \ToolPASDA{}, is also partition-based, but \emph{does} use UIFs and UIF refinement. Additionally, \ToolPASDA{} newly introduces best effort classification of undecided cases, which results in a classification of \scUnknown{} results into \scMaybeEq{}, \scMaybeNeq{}, and \scUnknown{}.

\begin{table}[tbp]
{
\caption{Properties of the evaluated equivalence checking approaches.}
\setlength{\tabcolsep}{5pt}
\begin{tabularx}{\columnwidth}{lXccc} \toprule
Tool          & Type            & UIFs & UIF        & \sc{Maybe} \\
              &                 &           & Refinement & Classes \\ \midrule
\ToolPASDA{}  & Partition-based & \cmark    & \cmark     & \cmark  \\
\ToolARDiff{} & Summary-based   & \cmark    & \cmark     & \xmark  \\
\ToolDSE{}    & Summary-based   & \cmark    & \xmark     & \xmark  \\
\ToolPRV{}    & Partition-based & \xmark    & \xmark     & \xmark  \\ \bottomrule
\end{tabularx}
\label{tab:properties-of-approaches}
}
\end{table}

\paragraph{\ToolARDiff{}} The publicly available implementation of \ToolARDiff{} \citep{badihi2020github} was modified by us to fix two bugs that caused false positive \scNeq{} results in the original \ToolARDiff{} implementation for a small number of cases. The first fix concerns the incorrect handling of temporary variables introduced by SPF to hold the results of explicit and implicit type casts between integers and doubles. The second fix concerns the incorrect handling of programs that are depth limited for different input partitions. We provide further details about the bugs and fixes in our replication package~\citep{replicationpackage}.

\paragraph{\ToolDSE{}} For \ToolDSE{}, we also used a fixed version of the tool that is based on a reimplementation by \citet{badihi2020github}. Use of the original implementation of \ToolDSE{} is not possible since the tool is not publicly available \citep{person2008differential,badihi2020ardiff}. The reimplementation by \citeauthor{badihi2020github} faithfully reconstructs the equivalence checking process described in the \ToolDSE{} paper but does not include the calculation of behavioral deltas. The fixes that we applied are the same as for \ToolARDiff{} and are also available in our replication package~\citep{replicationpackage}.

\paragraph{\ToolPRV{}} No publicly available implementation of \ToolPRV{} exists and we could not get access to the tool by contacting the authors of the \ToolPRV{} paper. %
We therefore used a variant of \ToolPASDA{} that exhibits the same basic properties as \ToolPRV{} (see Table~\ref{tab:properties-of-approaches}) as a proxy for \ToolPRV{} in our evaluation.
Specifically, this variant only performs \ToolPASDA{}'s first iteration and does not include best effort classification of undecided cases.
While the absolute classification accuracies and runtimes of this variant are likely different from the original \ToolPRV{} implementation, %
having this variant allows us to compare the relative advantages and disadvantages of \ToolPASDA{}'s extensions compared to a \ToolPRV{}-like baseline. %

\subsection{Benchmark Programs}
\label{sec:benchmark-programs}

For our evaluation, we used the equivalence checking benchmark built by \citet{badihi2020ardiff} to evaluate \ToolARDiff{}.
The benchmark consists of 141 Java program pairs (73 equivalent, 68 non-equivalent) with seeded changes.
All programs are single-threaded and deterministic, and every program is implemented in a single Java class consisting of one or more methods.
Program sizes range from 8 lines of code (LOC) to 201 LOC, with an average size of 52.5 LOC.
Program constructs used in the benchmark include loops, method calls, and non-linear arithmetic, but do not include recursion. Data types in the programs are limited to integers and doubles.
Other data types were not included by the authors of the benchmark because SPF and Z3 only offer limited support for non-primitive data types such as arrays, strings, and other classes.

\subsection{RQ1: Equivalence Classification Accuracy}
\label{sec:program-classification-results}

\begin{table*}[t]
{
\captionsetup{justification=centering}
\caption{Program-level \scEq{}, \scNeq{}, and \scUnknown{}\textquotesingle{} classifications of the four evaluated tools at the six evaluated timeout settings.\\To aid comparison across tools, \scUnknown{}\textquotesingle{} shows an aggregate of \scMaybeEq{}, \scMaybeNeq{}, and \scUnknown{} classifications for \ToolPASDA{}.}
\label{tab:program-level-classifications-1}
\setlength{\tabcolsep}{3pt}
\begin{tabularx}{\textwidth}{l@{\quad}l@{\quad}rcrcrcrcrcrXrcrcrcrcrcrXrcrcrcrcrcr} \toprule
Tool   & Expected & \multicolumn{11}{c}{\scEq{}}                          &  & \multicolumn{11}{c}{\scNeq{}}                         &  & \multicolumn{11}{c}{\scUnknown{}\textquotesingle}                   \\ \cmidrule{3-13} \cmidrule{15-25} \cmidrule{27-37}
       &          & \multicolumn{1}{c}{\rotatebox{90}{10~s}} & & \multicolumn{1}{c}{\rotatebox{90}{30~s}} & & \multicolumn{1}{c}{\rotatebox{90}{90~s}} & & \multicolumn{1}{c}{\rotatebox{90}{300~s}} & & \multicolumn{1}{c}{\rotatebox{90}{900~s}} & & \multicolumn{1}{c}{\rotatebox{90}{3600~s}} & & \multicolumn{1}{c}{\rotatebox{90}{10~s}} & & \multicolumn{1}{c}{\rotatebox{90}{30~s}} & & \multicolumn{1}{c}{\rotatebox{90}{90~s}} & & \multicolumn{1}{c}{\rotatebox{90}{300~s}} & & \multicolumn{1}{c}{\rotatebox{90}{900~s}} & & \multicolumn{1}{c}{\rotatebox{90}{3600~s}} & & \multicolumn{1}{c}{\rotatebox{90}{10~s}} & & \multicolumn{1}{c}{\rotatebox{90}{30~s}} & & \multicolumn{1}{c}{\rotatebox{90}{90~s}} & & \multicolumn{1}{c}{\rotatebox{90}{300~s}} & & \multicolumn{1}{c}{\rotatebox{90}{900~s}} & & \multicolumn{1}{c}{\rotatebox{90}{3600~s}} \\ \midrule
\ToolPASDA{}  & \scEq{}  & 38 &  & 41 &  & 45 &  & 47 &  & 50 &  & 51 &  & 0  &  & 0  &  & 0  &  & 0  &  & 0  &  & 0  &  & 32 &  & 31 &  & 27 &  & 25 &  & 23 &  & 22 \\
\ToolPASDA{}  & \scNeq{} & 0  &  & 0  &  & 0  &  & 0  &  & 0  &  & 0  &  & 48 &  & 50 &  & 50 &  & 52 &  & 52 &  & 53 &  & 17 &  & 17 &  & 17 &  & 15 &  & 16 &  & 14 \\ \midrule
\ToolARDiff{} & \scEq{}  & 45 &  & 48 &  & 48 &  & 48 &  & 48 &  & 49 &  & 0  &  & 0  &  & 0  &  & 0  &  & 0  &  & 0  &  & 16 &  & 16 &  & 18 &  & 19 &  & 19 &  & 18 \\
\ToolARDiff{} & \scNeq{} & 0  &  & 0  &  & 0  &  & 0  &  & 0  &  & 0  &  & 33 &  & 35 &  & 35 &  & 36 &  & 37 &  & 37 &  & 27 &  & 26 &  & 26 &  & 24 &  & 22 &  & 21 \\ \midrule
\ToolDSE{}    & \scEq{}  & 30 &  & 30 &  & 30 &  & 30 &  & 30 &  & 30 &  & 0  &  & 0  &  & 0  &  & 0  &  & 0  &  & 0  &  & 32 &  & 36 &  & 38 &  & 38 &  & 37 &  & 37 \\
\ToolDSE{}    & \scNeq{} & 0  &  & 0  &  & 0  &  & 0  &  & 0  &  & 0  &  & 21 &  & 21 &  & 21 &  & 21 &  & 21 &  & 21 &  & 39 &  & 40 &  & 40 &  & 40 &  & 40 &  & 41 \\ \midrule
\ToolSE{}     & \scEq{}  & 34 &  & 37 &  & 39 &  & 41 &  & 41 &  & 42 &  & 0  &  & 0  &  & 0  &  & 0  &  & 0  &  & 0  &  & 36 &  & 35 &  & 33 &  & 31 &  & 32 &  & 31 \\
\ToolSE{}     & \scNeq{} & 0  &  & 0  &  & 0  &  & 0  &  & 0  &  & 0  &  & 48 &  & 50 &  & 50 &  & 52 &  & 52 &  & 53 &  & 17 &  & 17 &  & 17 &  & 15 &  & 16 &  & 15 \\ \bottomrule
\end{tabularx}
}
\end{table*}

\begin{table*}[t]
{
\captionsetup{justification=centering}
\caption{Program-level \scDepthLimited{}, \scTimeout{}, and \scError{} classifications of the four evaluated tools at the six evaluated timeout settings.}
\label{tab:program-level-classifications-2}
\setlength{\tabcolsep}{3.6pt}
\begin{tabularx}{\textwidth}{l@{\quad}l@{\quad}rcrcrcrcrcrXrcrcrcrcrcrXrcrcrcrcrcr}  \toprule
Tool   & Expected & \multicolumn{11}{c}{\scDepthLimited{}}         &  & \multicolumn{11}{c}{\scTimeout{}}                 &  & \multicolumn{11}{c}{\scError{}}                  \\ \cmidrule{3-13} \cmidrule{15-25} \cmidrule{27-37}
       &          & \multicolumn{1}{c}{\rotatebox{90}{10~s}} & & \multicolumn{1}{c}{\rotatebox{90}{30~s}} & & \multicolumn{1}{c}{\rotatebox{90}{90~s}} & & \multicolumn{1}{c}{\rotatebox{90}{300~s}} & & \multicolumn{1}{c}{\rotatebox{90}{900~s}} & & \multicolumn{1}{c}{\rotatebox{90}{3600~s}} & & \multicolumn{1}{c}{\rotatebox{90}{10~s}} & & \multicolumn{1}{c}{\rotatebox{90}{30~s}} & & \multicolumn{1}{c}{\rotatebox{90}{90~s}} & & \multicolumn{1}{c}{\rotatebox{90}{300~s}} & & \multicolumn{1}{c}{\rotatebox{90}{900~s}} & & \multicolumn{1}{c}{\rotatebox{90}{3600~s}} & & \multicolumn{1}{c}{\rotatebox{90}{10~s}} & & \multicolumn{1}{c}{\rotatebox{90}{30~s}} & & \multicolumn{1}{c}{\rotatebox{90}{90~s}} & & \multicolumn{1}{c}{\rotatebox{90}{300~s}} & & \multicolumn{1}{c}{\rotatebox{90}{900~s}} & & \multicolumn{1}{c}{\rotatebox{90}{3600~s}} \\ \midrule
\ToolPASDA{}  & \scEq{}  & 1 &  & 0 &  & 0 &  & 1 &  & 0 &  & 0 &  & 2  &  & 1 &  & 1 &  & 0 &  & 0 &  & 0 &  & 0 &  & 0 &  & 0 &  & 0 &  & 0 &  & 0 \\
\ToolPASDA{}  & \scNeq{} & 1 &  & 0 &  & 0 &  & 1 &  & 0 &  & 0 &  & 2  &  & 1 &  & 1 &  & 0 &  & 0 &  & 0 &  & 0 &  & 0 &  & 0 &  & 0 &  & 0 &  & 1 \\ \bottomrule
\ToolARDiff{} & \scEq{}  & 0 &  & 0 &  & 0 &  & 0 &  & 0 &  & 0 &  & 12 &  & 9 &  & 7 &  & 6 &  & 4 &  & 4 &  & 0 &  & 0 &  & 0 &  & 0 &  & 2 &  & 2 \\
\ToolARDiff{} & \scNeq{} & 0 &  & 0 &  & 0 &  & 0 &  & 0 &  & 0 &  & 8  &  & 7 &  & 7 &  & 7 &  & 6 &  & 5 &  & 0 &  & 0 &  & 0 &  & 1 &  & 3 &  & 5 \\ \bottomrule
\ToolDSE{}    & \scEq{}  & 0 &  & 0 &  & 0 &  & 0 &  & 0 &  & 0 &  & 11 &  & 7 &  & 5 &  & 5 &  & 5 &  & 4 &  & 0 &  & 0 &  & 0 &  & 0 &  & 1 &  & 2 \\
\ToolDSE{}    & \scNeq{} & 0 &  & 0 &  & 0 &  & 0 &  & 0 &  & 0 &  & 8  &  & 7 &  & 7 &  & 7 &  & 6 &  & 5 &  & 0 &  & 0 &  & 0 &  & 0 &  & 1 &  & 1 \\ \bottomrule
\ToolSE{}     & \scEq{}  & 1 &  & 0 &  & 0 &  & 1 &  & 0 &  & 0 &  & 2  &  & 1 &  & 1 &  & 0 &  & 0 &  & 0 &  & 0 &  & 0 &  & 0 &  & 0 &  & 0 &  & 0 \\
\ToolSE{}     & \scNeq{} & 1 &  & 0 &  & 0 &  & 1 &  & 0 &  & 0 &  & 2  &  & 1 &  & 1 &  & 0 &  & 0 &  & 0 &  & 0 &  & 0 &  & 0 &  & 0 &  & 0 &  & 0 \\ \bottomrule
\end{tabularx}
}
\end{table*}

\paragraph{Setup} We collected the equivalence classification results for all 141 cases in the \ToolARDiff{} benchmark for each of the four evaluated tools at six different timeout settings: 10~seconds~(s), 30~s, 90~s, 300~s, 900~s, and 3600~s. To further improve confidence in the collected results, we conducted five runs for each benchmark:tool:timeout combination, resulting in a total of 16920 runs across the 3384 five-run-groups. For 3360 of the five-run-groups, all five runs produced the same classification result. For the remaining 24 groups -- all of which produced two different results across the five runs in the group -- we report the classification result produced by the majority of the runs as the group result. Inter-group classification differences are generally caused by slight runtime differences across the runs of a group, which affect the analysis progress that can be made within the given timeout. Thus, inter-group differences can be observed across all tools and all timeout settings.

All experiments were run on a 2022 MacBook Air with M2 chip and 24~GB of RAM. Java's \texttt{InitialHeapSize} and \texttt{MaxHeapSize} settings were left at their default values of 384~MB (i.e., ${1/64}$th of RAM) and 6~GB (i.e., ${1/4}$th of RAM), respectively. The four tools were configured to use a depth limit setting of 10, which we found to offer a good trade-off between tool runtimes and result accuracy in our own testing. Collected results were stored in an SQLite database and analyzed via SQL queries to produce the results of the evaluation. All collected data and the corresponding analysis scripts are available in our replication package~\citep{replicationpackage}.

\paragraph{Results}
Tables~\ref{tab:program-level-classifications-1} and \ref{tab:program-level-classifications-2} show the program-level classification results of the four tools. Overall, \ToolPASDA{} correctly classifies between 86 of 141 cases (61\%) at the 10~s timeout and 104 of 141 cases (74\%) at the 3600~s timeout, with correctly classified cases at higher timeouts being (proper) supersets of those that are correctly classified at lower timeouts. Looking at equivalent and non-equivalent cases separately, we find that \ToolPASDA{} correctly classifies 38 of 73 (52\%) equivalent and 48 of 68 (61\%) non-equivalent cases when using a timeout setting of 10~s. This increases to 51 of 73 (70\%) equivalent and 53 of 68 (78\%) non-equivalent cases at the 3600~s timeout setting. Most remaining cases are classified as \scUnknown{}\textquotesingle{} (i.e., \scMaybeEq{}, \scMaybeNeq{} or \scUnknown{} --- for further details, see Section~\ref{sec:best-effort-classification-results}). 
Only a small minority of cases are classified as \scDepthLimited{}, \scTimeout{} or \scError{}, with all \scError{} classifications across all tools being caused exclusively by \texttt{OutOfMemoryError}s. No \scEq{} cases are incorrectly classified as \scNeq{} or vice versa. 

Compared to \ToolSE{}, \ToolPASDA{} correctly classifies a higher number of \scEq{} cases (e.g., 38 vs.\ 34 at 10~s and 51 vs.\ 42 at 3600~s) and the same number of \scNeq{} cases. In fact, the set of cases that is correctly classified by \ToolPASDA{} is a (proper) superset of those that are correctly classified by \ToolSE{}. This is generally guaranteed by our experimental setup (\ToolSE{} is identical to the first iteration of \ToolPASDA{}). Nevertheless, these results demonstrate that the use of UIFs and UIF refinement can improve the classification accuracy of \scEq{} cases for not only summary-based tools -- as demonstrated by \ToolARDiff{} \citep{badihi2020ardiff} -- but also for partition-based tools such as \ToolPASDA{}, albeit at the cost of longer runtimes (see Section~\ref{sec:runtime-performance-results}). \scNeq{} cases, on the other hand, generally do not benefit from abstraction via UIFs because it is usually not possible to tell whether non-equivalences that were identified in the abstracted programs (which overapproximate the original program behaviors) can also be observed in the original programs.

Compared to \ToolARDiff{}, \ToolPASDA{} correctly classifies more cases at all timeout settings (e.g., 86 vs.\ 78 at 10~s and 104 vs.\ 86 at 3600~s). For \scEq{} cases, \ToolPASDA{} correctly classifies fewer cases than \ToolARDiff{} at low timeouts (e.g., 38 vs.\ 45 at 10~s) and more at high timeouts (e.g., 51 vs.\ 49 at 3600~s). For \scNeq{} cases, \ToolPASDA{} correctly classifies more cases than \ToolARDiff{} at all timeout settings (e.g., 48 vs.\ 33 at 10~s and 53 vs.\ 37 at 3600~s). While there is significant overlap across the sets of cases that are correctly classified by \ToolPASDA{} and \ToolARDiff{}, neither is a superset of the other. For example, at the 300~s timeout, 44 of the 73 \scEq{} cases are correctly classified by both tools, whereas 3 are only correctly classified by \ToolPASDA{} and 4 by \ToolARDiff{}. At the same 300~s timeout, 35 of the 68  \scNeq{} cases are correctly classified by both tools, 17 only by \ToolPASDA{}, and 1 only by \ToolARDiff{}. \ToolPASDA{} generally performs better than \ToolARDiff{} on cases that are best analyzed without abstraction, i.e., without using UIFs to represent unchanged parts of the source code. This is because \ToolPASDA{} checks equivalence without abstraction in its first iteration, whereas \ToolARDiff{} does this in its last iteration (which generally is not reached within the given timeout). On the other hand, \ToolPASDA{} generally performs worse than \ToolARDiff{} for \scEq{} cases that \ToolARDiff{} \emph{can} fully analyze within the given timeout but \ToolPASDA{} \emph{cannot} due to its higher runtime requirements (see Section~\ref{sec:runtime-performance-results} for further details about runtime differences). %

Compared to \ToolDSE{}, which is identical to \ToolARDiff{}'s first iteration, \ToolPASDA{} correctly classifies more \scEq{} cases as well as more \scNeq{} cases at all timeout settings. Intuitively, the comparatively low classification accuracy of \ToolDSE{} can be attributed to the fact that \ToolDSE{} introduces UIFs not only to replace complex program constructs that are difficult to reason about (e.g., loops, non-linear arithmetic, etc.) but also for simple ones (e.g., variable assignments, linear arithmetic, etc.). This results in unnecessary overapproximations that hinder non-/equivalence proofs. \ToolPASDA{} and \ToolARDiff{} mitigate these negative effects through their abstraction refinement process as proposed by \citet{badihi2020ardiff}. However, this comes at the cost of longer runtimes (see Section~\ref{sec:runtime-performance-results}) because multiple analysis iterations have to be performed to find an appropriate level of abstraction.

\medbreak
\noindent\fbox{\begin{minipage}{0.97\columnwidth}
\textbf{Answer to RQ1}: \ToolPASDA{} achieves an overall program-level equivalence classification accuracy between 61\% at the 10~s timeout setting and 74\% at the 3600~s timeout setting. Compared to the three existing tools included in our analysis, which achieve classification accuracies between 36--58\% at the 10~s timeout setting and 36--67\% at the 3600~s timeout setting, \ToolPASDA{} therefore correctly classifies a larger number of cases at all six of the analyzed timeout settings.
\end{minipage}}

\subsection{RQ2: Best Effort Classification Accuracy}

\label{sec:best-effort-classification-results}
\begin{table*}[b]
{
\captionsetup{justification=centering}
\caption{Program-level \scMaybeEq{}, \scMaybeNeq{}, and \scUnknown{} classifications at the six evaluated timeout settings.\\ \ToolARDiff{}, \ToolDSE{}, and \ToolPRV{} were modified by us to differentiate between these three classes for cases that were previously classified as \scUnknown{}.}
\label{tab:program-level-classifications-3}
\setlength{\tabcolsep}{3.3pt}
\begin{tabularx}{\textwidth}{l@{\quad}l@{\quad}rcrcrcrcrcrXrcrcrcrcrcrXrcrcrcrcrcr}  \toprule
Tool   & Expected & \multicolumn{11}{c}{\scMaybeEq{}}         &  & \multicolumn{11}{c}{\scMaybeNeq{}}                 &  & \multicolumn{11}{c}{\scUnknown{}}                  \\ \cmidrule{3-13} \cmidrule{15-25} \cmidrule{27-37}
       &          & \multicolumn{1}{c}{\rotatebox{90}{10~s}} & & \multicolumn{1}{c}{\rotatebox{90}{30~s}} & & \multicolumn{1}{c}{\rotatebox{90}{90~s}} & & \multicolumn{1}{c}{\rotatebox{90}{300~s}} & & \multicolumn{1}{c}{\rotatebox{90}{900~s}} & & \multicolumn{1}{c}{\rotatebox{90}{3600~s}} & & \multicolumn{1}{c}{\rotatebox{90}{10~s}} & & \multicolumn{1}{c}{\rotatebox{90}{30~s}} & & \multicolumn{1}{c}{\rotatebox{90}{90~s}} & & \multicolumn{1}{c}{\rotatebox{90}{300~s}} & & \multicolumn{1}{c}{\rotatebox{90}{900~s}} & & \multicolumn{1}{c}{\rotatebox{90}{3600~s}} & & \multicolumn{1}{c}{\rotatebox{90}{10~s}} & & \multicolumn{1}{c}{\rotatebox{90}{30~s}} & & \multicolumn{1}{c}{\rotatebox{90}{90~s}} & & \multicolumn{1}{c}{\rotatebox{90}{300~s}} & & \multicolumn{1}{c}{\rotatebox{90}{900~s}} & & \multicolumn{1}{c}{\rotatebox{90}{3600~s}} \\ \midrule
\ToolPASDA{}  & \scEq{}  & 28 &  & 28 &  & 23 &  & 19 &  & 20 &  & 19 &  & 2  &  & 2  &  & 3  &  & 4  &  & 1  &  & 1  &  & 2 &  & 1  &  & 1  &  & 2  &  & 2  &  & 2  \\
\ToolPASDA{}  & \scNeq{} & 10 &  & 9  &  & 8  &  & 7  &  & 7  &  & 8  &  & 4  &  & 4  &  & 5  &  & 5  &  & 5  &  & 4  &  & 3 &  & 4  &  & 4  &  & 3  &  & 4  &  & 2  \\ 
\midrule
\ToolARDiff{} & \scEq{}  & 8  &  & 9  &  & 10 &  & 11 &  & 11 &  & 11 &  & 5  &  & 1  &  & 1  &  & 1  &  & 0  &  & 0  &  & 3 &  & 6  &  & 7  &  & 7  &  & 8  &  & 7  \\
\ToolARDiff{} & \scNeq{} & 2  &  & 2  &  & 2  &  & 2  &  & 2  &  & 2  &  & 16 &  & 10 &  & 10 &  & 8  &  & 7  &  & 6  &  & 9 &  & 14 &  & 14 &  & 14 &  & 13 &  & 13 \\ \midrule
\ToolDSE{}    & \scEq{}  & 5  &  & 6  &  & 6  &  & 6  &  & 6  &  & 6  &  & 19 &  & 20 &  & 21 &  & 21 &  & 21 &  & 21 &  & 8 &  & 10 &  & 11 &  & 11 &  & 10 &  & 10 \\
\ToolDSE{}    & \scNeq{} & 2  &  & 2  &  & 2  &  & 2  &  & 2  &  & 2  &  & 32 &  & 33 &  & 33 &  & 33 &  & 34 &  & 35 &  & 5 &  & 5  &  & 5  &  & 5  &  & 4  &  & 4  \\ \midrule
\ToolSE{}     & \scEq{}  & 33 &  & 32 &  & 30 &  & 28 &  & 29 &  & 28 &  & 0  &  & 0  &  & 0  &  & 0  &  & 0  &  & 0  &  & 3 &  & 3  &  & 3  &  & 3  &  & 3  &  & 3  \\
\ToolSE{}     & \scNeq{} & 11 &  & 10 &  & 9  &  & 8  &  & 8  &  & 8  &  & 2  &  & 2  &  & 3  &  & 3  &  & 3  &  & 3  &  & 4 &  & 5  &  & 5  &  & 4  &  & 5  &  & 4 \\ 
\bottomrule
\end{tabularx}
}
\end{table*}

\paragraph{Setup} To evaluate the accuracy of \ToolPASDA{}'s best effort equivalence classifications, we used the same basic setup as for RQ1. For the purpose of accuracy calculations, we consider \scMaybeEq{} classifications as correct if they are reported for \scEq{} benchmark cases, and as incorrect if they are reported for \scNeq{} benchmark cases. Similarly, we consider \scMaybeNeq{} classifications as correct when reported for \scNeq{} cases and as incorrect when reported for \scEq{} cases. To allow us to compare the effectiveness of \ToolPASDA{}'s best effort classification approach across different baselines, we retrofitted it to the three existing tools \ToolARDiff{}, \ToolDSE{}, and \ToolPRV{}. This causes a classification of their \scUnknown{} results into \scMaybeEq{}, \scMaybeNeq{} and \scUnknown{}. None of their other classifications are affected by this change. %

\paragraph{Results} Table~\ref{tab:program-level-classifications-3} shows the best effort classification results for all benchmark cases that were previously classified as \scUnknown{} (see columns \scUnknown{}\textquotesingle{} in Table~\ref{tab:program-level-classifications-2}). For \ToolPASDA{}, around 10--15\% of these cases retain their \scUnknown{} classifications (e.g., 5 of 49 at the 10~s timeout and 4 of 36 at 3600~s), signifying that none of the fully analyzed partitions provide any indication of potential non-/equivalence. Around 65--80\% of cases (e.g., 38 of 49 at 10~s and 27 of 36 at 3600~s) are classified as \scMaybeEq{} and 10--25\% are classified as \scMaybeNeq{} (e.g., 6 of 49 at 10~s and 5 of 36 at 3600~s).

Among the \scMaybeEq{} cases, around 70--75\% are correctly classified (e.g., 28 of 38 at 10~s and 19 of 27 at 3600~s), whereas the remaining 25--30\% are incorrectly classified. Incorrect \scMaybeEq{} classifications arise for the following two reasons: (i)~the programs were only partially analyzed (due to the configured timeout and/or depth limit) and the non-equivalent partitions were \emph{not} reached by the analysis, (ii)~the non-equivalent partitions \emph{were} reached by the analysis but non-equivalence could not be proven due to limitations of Z3, which we use for non-/equivalence checks. These reasons directly follow from the definitions of our classification heuristics (see Sections~\ref{sec:partition-equivalence-classification} and \ref{sec:program-equivalence-classification}) and could thus potentially be improved upon through the use of a different set of heuristics. We leave such refinements for future research.

Among the \scMaybeNeq{} cases, fluctuations in classification accuracy are higher than for \scMaybeEq{} cases, with correct classifications reaching 55--85\% across the different timeouts (e.g., 4 of 6 at 10~s and 4 of 5 at 3600~s). Incorrect \scMaybeNeq{} classifications primarily arise when (i)~the observed non-equivalent program behaviors only manifest due the overapproximations caused by the introduction of UIFs or (ii)~partitions with a reachability classification of \scMaybeReachable{} and an output classification of \scNeq{} are actually \scUnreachable{} but could not be proven to be so due to limitations of Z3, which we use for reachability checks. Again, classifications for such cases could potentially be improved upon through the use of a different set of classification heuristics. Alternatively, the concrete input values that are provided by \ToolPASDA{} to demonstrate non-equivalence could be used to check whether concrete executions of the two original program versions produce non-equivalent results for these inputs. If they do, the result is a true positive. Otherwise, it is a false positive. Note that this approach \emph{cannot} be used to check the correctness of \scMaybeEq{} classifications. This is because, in the \scMaybeEq{} case, we need to prove that the two programs are equivalent for \emph{all} inputs, which cannot be achieved by checking a single set of input values.

Best effort classification accuracies show similar trends for \ToolARDiff{}, \ToolDSE{}, and \ToolSE{} as they do for \ToolPASDA{}. Specifically,  correct classifications are consistently above 60\% for both \scMaybeEq{} and \scMaybeNeq{} cases at all timeout settings. For high timeout settings of \ToolARDiff{} and all timeout settings of \ToolSE{}, classification accuracies of \scMaybeNeq{} cases even reach 100\%. However, this is not guaranteed in general, since \ToolSE{} can still produce false positive \scMaybeNeq{} classifications in the presence of incorrect \scMaybeReachable{} reachability classifications, and \ToolARDiff{} can additionally produce false positive \scMaybeNeq{} classifications due to overapproximations introduced by UIFs. \ToolARDiff{} and \ToolDSE{} produce more \scMaybeNeq{} results than \ToolPASDA{} and \ToolSE{} because the two partition-based tools can often prove the corresponding cases to be \scNeq{} instead. Similarly, \ToolPASDA{} produces fewer \scMaybeEq{} results than \ToolSE{} because it can often prove them to be \scEq{} instead.

\begin{table*}[t]
{
\captionsetup{justification=centering}
\caption{Occurrence frequencies of different partition-level classifications when running \ToolPASDA{} with timeout setting of 300~s.\\Partition-level classifications (columns 4--9) that cannot occur for the corresponding program-level classification (column 3) are marked with a ``-".}%
\label{tab:partition-level-classifications}
\setlength{\tabcolsep}{9pt}
\begin{tabularx}{\textwidth}{rlXrrrrrrrr} \toprule
\# & Program-level & Program-level & \multicolumn{6}{c}{Average \% of \textless{}Classification\textgreater{} Partitions per Run} \\ \cmidrule{4-9}
   & Expected      & Actual        & \scEq{} & \scNeq{} & \scMaybeEq{} & \scMaybeNeq{} & \scUnknown{} & \scDepthLimited{} \\ \midrule
1  & \scEq{}  & \scEq{}           & 100.0 &    - & -   &    - &     - &     - \\
2  & \scEq{}  & \scMaybeEq{}      &  67.3 &    - & 0.0 &    - &   0.9 &  31.8 \\
3  & \scEq{}  & \scMaybeNeq{}     &  54.8 &    - & 1.9 & 21.9 &   1.9 &  19.6 \\
4  & \scEq{}  & \scUnknown{}      &     - &    - & -   &    - & 100.0 &   0.0 \\
5  & \scEq{}  & \scDepthLimited{} &     - &    - & -   &    - &     - & 100.0 \\ \midrule
6  & \scNeq{} & \scNeq{}          &  22.3 & 42.5 & 0.0 &  0.9 &   7.3 &  27.0 \\
7  & \scNeq{} & \scMaybeEq{}      &  58.4 &    - & 0.0 &    - &  35.1 &   6.6 \\
8  & \scNeq{} & \scMaybeNeq{}     &  35.9 &    - & 0.0 & 33.7 &   1.5 &  28.9 \\
9  & \scNeq{} & \scUnknown{}      &     - &    - & -   &    - &  60.0 &  40.0 \\
10 & \scNeq{} & \scDepthLimited{} &     - &    - & -   &    - &     - & 100.0 \\ \midrule
11 & *        & *                 &  50.8 & 20.5 & 0.1 &  3.5 &   6.1 &  18.9 \\ \bottomrule
\end{tabularx}
}
\end{table*}

As described in Section~\ref{sec:program-equivalence-classification}, \emph{program-level} (best effort) classifications are generally derived from \emph{partition-level} (best effort) classifications through a set of classification heuristics. Exceptions to this rule are (i) runs during which an error occurs, which are always classified as \scError{}, and (ii) runs for which not even a single partition can be analyzed within the given timeout, which are always classified as \scTimeout{}. For the retrofitted variants of \ToolARDiff{} and \ToolDSE{}, we consider the single program-level equivalence check to be representative of a partition that covers the whole input space. Otherwise, the same classification heuristics are applied.

To provide a better understanding of these heuristics, Table~\ref{tab:partition-level-classifications} shows how often each partition-level result is observed, on average, across runs with a particular program-level classification when running \ToolPASDA{} with a timeout setting of 300~s. By definition, a program-level result of \scEq{} is only reported when \ToolPASDA{} is able to analyze all partitions within the given timeout and proves them all to be \scEq{} (see row \#1). \scMaybeEq{} is reported when some partitions are classified as \scEq{} or \scMaybeEq{}, but none are classified as \scNeq{} or \scMaybeNeq{} (see rows \#2 and \#7). Otherwise, cases are classified based on the partition with the highest priority, where \scNeq{} \textgreater{} \scMaybeNeq{} \textgreater{} \scUnknown{} \textgreater{} \scDepthLimited{}. Thus, program-level \scNeq{} cases are the only ones that can contain \scNeq{} partitions, but might also contain all other types of partitions (see row \#6). Similarly, \scNeq{} and \scMaybeNeq{} cases are the only ones that can contain \scMaybeNeq{} partitions (see rows \#3, \#6, and \#8). \scDepthLimited{} cases, on the other hand, can only contain \scDepthLimited{} partitions (see rows \#5 and \#10), since any other partition-level result would force a different program-level classification to be chosen. For the same reason, \scUnknown{} cases can only contain \scUnknown{} and \scDepthLimited{} partitions (see rows \#4 and \#9).

\medbreak
\noindent\fbox{\begin{minipage}{0.97\columnwidth}
\textbf{Answer to RQ2}: 
Through the introduction of our best effort classification approach, 65-80\% of previously \scUnknown{} cases are classified as \scMaybeEq{}, 10--25\% as \scMaybeNeq{}, and 10--15\% retain their \scUnknown{} classifications. \scMaybeEq{}s are correct for 70--75\% of cases across the six evaluated timeout settings between 10~s and 3600~s. \scMaybeNeq{}s are correct for 55--85\% of cases. We observed similar best effort classification accuracies when retrofitting \ToolPASDA{}'s best effort classification approach to \ToolARDiff{}, \ToolDSE{} and \ToolPRV{}.
\end{minipage}}

\subsection{RQ3: Partition-Level Non-/Equivalence Proofs}
\label{sec:partition-classification-results}

\paragraph{Setup} 
To answer RQ3, we executed \ToolPASDA{} and \ToolPRV{} five times each for every case in the benchmark across each of the six timeout settings of 10~s, 30~s, 90~s, 300~s, 900~s, and 3600~s. \ToolARDiff{} and \ToolDSE{} are not included in the partition-level analysis since neither tool produces partition-level results. The remaining hardware and tool configurations are the same as for RQ1.

\paragraph{Results}
Table~\ref{tab:overall-partition-level-classifications} provides an overview of the partition-level equivalence classification results produced by \ToolPASDA{} and \ToolPRV{} across the six analyzed timeout settings. In total, \ToolPASDA{} identifies 3298 partitions across the 705 runs (i.e., 5 runs for each of the 141 benchmark cases) at the 10~s timeout setting. This number increases to 5535 partitions at the 3600~s timeout setting, whereas \ToolPRV{} identifies between 3500 partitions at 10~s and 5809 at 3600~s. The number of reported partitions is generally lower for \ToolPASDA{} than for \ToolPRV{} because of \ToolPASDA{}'s iterative abstraction and refinement process. After all, partition counts can decrease if the data of the last (partially analyzed) iteration is reported as \ToolPASDA{}'s analysis result (see Section~\ref{sec:program-output}), and partition counts can further fluctuate as UIFs are added and removed throughout \ToolPASDA{}'s analysis process. This also explains why \ToolPASDA{} reports fewer partitions when raising its timeout setting from 300~s to 900~s.

Partition counts across different benchmark cases are highly skewed. For example, partition counts reported by \ToolPASDA{} at the 300~s timeout setting have a skewness value of 1.2 (mean: 7.8, median: 4, mode: 3), which indicates that most cases in the benchmark have relatively few partitions, but some cases exist that have much higher partition counts. High partition counts are most commonly found among cases that contain (nested) loops and/or many (nested) if constructs. Furthermore, \scNeq{} cases tend to have higher partition counts than corresponding \scEq{} cases. Intuitively, this is because the introduction of semantic changes often causes some \scEq{} partitions to be split into multiple \scEq{} and \scNeq{} subpartitions. %

\begin{table*}[t]
{
\captionsetup{justification=centering}
\caption{Overall partition-level equivalence classification results of \ToolPASDA{} and \ToolPRV{} across the six analyzed timeout settings.}
\label{tab:overall-partition-level-classifications}
\setlength{\tabcolsep}{9.5pt}
\begin{tabularx}{\textwidth}{lrrrrrrrrrr} \toprule
Tool & Timeout (s) & \# Partitions & \multicolumn{6}{c}{Average \% of \textless{}Classification\textgreater{} Partitions} \\ \cmidrule{4-9}
 & & & \scEq{} & \scNeq{} & \scMaybeEq{} & \scMaybeNeq{} & \scUnknown{} & \scDepthLimited{} \\ \midrule
\ToolPASDA{} & 10   & 3298 & 52.5 & 23.1 & 0.0 & 2.8 & 5.3 & 16.3 \\
\ToolPASDA{} & 30   & 4724 & 47.8 & 20.7 & 0.1 & 2.1 & 6.0 & 23.2 \\
\ToolPASDA{} & 90   & 5155 & 52.6 & 20.0 & 0.1 & 3.2 & 6.1 & 18.0 \\
\ToolPASDA{} & 300  & 5502 & 50.8 & 20.5 & 0.1 & 3.5 & 6.1 & 18.9 \\
\ToolPASDA{} & 900  & 5430 & 52.3 & 20.9 & 0.1 & 1.3 & 6.1 & 19.3 \\
\ToolPASDA{} & 3600 & 5535 & 51.4 & 20.4 & 0.1 & 1.5 & 5.7 & 20.9 \\ \midrule
\ToolPRV{}   & 10   & 3500 & 51.4 & 21.8 &  -  &  -  & 7.8 & 19.0 \\
\ToolPRV{}   & 30   & 4804 & 47.8 & 20.4 &  -  &  -  & 6.9 & 24.9 \\
\ToolPRV{}   & 90   & 5361 & 49.6 & 19.2 &  -  &  -  & 6.7 & 24.5 \\
\ToolPRV{}   & 300  & 5727 & 47.8 & 19.7 &  -  &  -  & 6.5 & 26.1 \\
\ToolPRV{}   & 900  & 5768 & 47.7 & 19.6 &  -  &  -  & 6.4 & 26.3 \\
\ToolPRV{}   & 3600 & 5809 & 47.4 & 19.7 &  -  &  -  & 6.4 & 26.5 \\ \bottomrule
\end{tabularx}
}
\end{table*}

Looking at \ToolPASDA{}'s classification results, we find that it proves 50.8\% of identified partitions to be \scEq{} and 20.5\% to be \scNeq{} at the 300~s timeout setting. For the remaining 28.7\% of \scMaybeEq{} (0.1\%), \scMaybeNeq{} (3.5\%), \scUnknown{} (6.1\%), and \scDepthLimited{} (18.9\%) partitions, no non-/equivalence proofs can be provided due to (i)~depth limiting, (ii)~UIF-induced overapproximations, and (iii)~limitations of Z3, which we use for reachability and non-/equivalence checks. As timeouts are changed, the proportions held by the different classifications fluctuate by a few percentage points, but no clear upward or downward trend emerges for any of them. However, since a larger number of partitions is analyzed at higher timeouts, this means that a larger part of the overall input space can be proven to be \scEq{} or \scNeq{}, or classified as either \scMaybeEq{} or \scMaybeNeq{}. 
For example, only 452 of 705 runs (64\%) finish at least one analysis iteration (i.e., fully analyze the original programs without UIFs) at the 10~s timeout, but 620 (88\%) do so at 3600~s.

Compared to \ToolPRV{}, \ToolPASDA{} provides (best effort) non-/equivalence classifications for a larger percentage of identified partitions across all six analyzed timeout settings. More specifically, \ToolPASDA{} classifies 0.0--4.6\% more partitions as \scEq{}, 0.3--1.3\% more as \scNeq{}, 0.0--0.1\% as \scMaybeEq{}, and 1.5--3.2\% as \scMaybeNeq{}. Consequently, the percentage of partitions for which no (best-effort) non-/equivalence classifications can be provided by \ToolPASDA{} is consistently lower than that of \ToolPRV{}, reaching a 0.4--2.5\% lower percentage of \scUnknown{} partitions and a 1.7--7.2\% lower percentage of \scDepthLimited{} partitions. These results match our expectations. After all, the introduction of UIFs intends to enable \scEq{} proofs for previously \scUnknown{} and \scDepthLimited{} partitions, but can also lead to an increase of other non-\scDepthLimited{} classifications if no \scEq{} proof can be provided. Additionally, the introduction of best effort classifications intends to replace some previously \scUnknown{} results with \scMaybeEq{} or \scMaybeNeq{}.

\medbreak
\noindent\fbox{\begin{minipage}{0.97\columnwidth}
\textbf{Answer to RQ3}: Across the six analyzed timeout settings, \ToolPASDA{} provides non-/equivalence \emph{proofs} for 70.7--78.4\% of identified partitions, whereas \ToolPRV{} provides such proofs for 67.1--73.2\% of identified partitions. At identical timeout settings, \ToolPASDA{} therefore provides non-/equivalence proofs for 2.5--7.4\% more of the total partitions than \ToolPRV{}. Additionally, \ToolPASDA{} provides \emph{best effort} non-/equivalence classifications for 1.4--3.6\% of identified partitions, whereas \ToolPRV{}, by design, cannot provide any best effort classifications.
\end{minipage}}

\subsection{RQ4: Runtime Performance}
\label{sec:runtime-performance-results}

\begin{figure*}[t!]
    \centering
    \includegraphics[keepaspectratio, width=\textwidth]{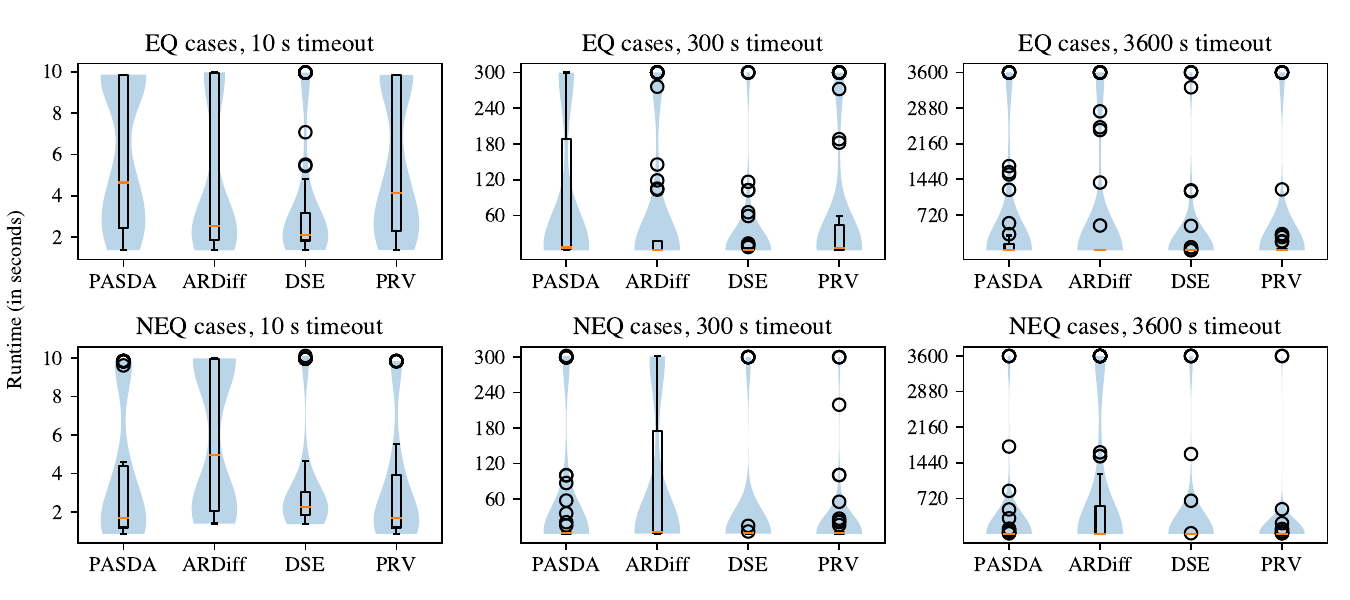}
    \captionsetup{justification=centering}
    \caption{
        Runtime distributions of the four evaluated tools (\ToolPASDA{}, \ToolARDiff{}, \ToolDSE{}, \ToolSE{}) \\ at timeout settings of 10~s, 300~s, and 3600~s per type of benchmark case (\scEq{} vs. \scNeq{}).%
    }
    \label{fig:runtime-plots-all}
\end{figure*}

\paragraph{Setup}
To answer RQ4, we reused the basic setup from RQ1 described before.
Thus, each of the four tools was executed five times for each benchmark case at each of the six analyzed timeout settings from 10~s to 3600~s.
Runtime measurements were taken for each processing step of the four tools using a custom wrapper around the Apache Commons \texttt{StopWatch} class~\citep{apachestopwatch}.
On average, there is around a 10\% difference between the fastest and the slowest run of each five-run-group.
We present the runtime results based on the median runtimes of each group to mitigate the influence of occasional random outliers.
The full runtime data is available in our replication package~\citep{replicationpackage}.

\paragraph{Results} Figure~\ref{fig:runtime-plots-all} shows the runtime distributions of the four analyzed tools at timeout settings of 10~s, 300~s, and 3600~s. Note that the measured runtimes are highly skewed for all tools, reaching skewness values between 0.91 and 2.00 across the different tool:timeout combinations. Consequently, median runtimes remain largely unchanged across the different timeouts, ranging from 2.2~s to 4.8~s across all tools. Mean runtimes, on the other hand, increase significantly as timeouts are raised. More specifically, they start at around 5~s for all tools at the 10~s timeout setting and go up to 300--700~s across the different tools at the 3600~s setting. %

For \scEq{} cases, \ToolDSE{} generally has the shortest runtimes of all tools because it only performs a single analysis iteration (whereas \ToolPASDA{} and \ToolARDiff{} perform multiple iterations) and often abstracts program constructs such as loops that would otherwise be expensive to analyze (whereas \ToolSE{} does not use any abstraction). For \scNeq{} cases, \ToolSE{} generally has the shortest runtimes because it only performs a single iteration (whereas \ToolPASDA{} performs multiple iterations) and can provide non-equivalence proofs without fully analyzing the two target programs (whereas \ToolARDiff{} and \ToolDSE{} have to fully analyze the target programs). \ToolPASDA{} has the \emph{longest} runtimes for \scEq{} cases and \ToolARDiff{} for \scNeq{} cases.

On average, \ToolPASDA{} takes around 6-17\% longer to analyze the same cases than \ToolSE{}.
This holds for both \scEq{} as well as \scNeq{} benchmark cases at all timeout settings, though the difference in runtimes is smallest at low timeouts and largest at high timeouts.
Since \ToolSE{} is identical to \ToolPASDA{}'s first iteration in our setup, runtime differences can be attributed to the iterative abstraction refinement process used by \ToolPASDA{}.
At the 10~s timeout setting, \ToolPASDA{} performs more than one iteration for 18\% of \scEq{} and 9\% of \scNeq{} cases.
These numbers go up as timeouts are increased because fewer cases are interrupted by the timeout before the analysis reaches later iterations,
which also explains the larger runtime difference between \ToolPRV{} and \ToolPASDA{} at higher timeouts.
At the 3600~s timeout setting, \ToolPASDA{} performs more than one iteration for 32\% of \scEq{} cases and 19\% of \scNeq{} cases.
The number of cases for which \ToolPASDA{} runs into the timeout decreases from 42\% of cases at 10~s to 18\% at 3600~s.

Compared to \ToolARDiff{}, \ToolPASDA{} is generally slower to provide equivalence proofs but faster to provide non-equivalence proofs.
More specifically, for \scEq{} cases that both tools classify correctly, \ToolPASDA{} takes, on average, 20--35\% longer to produce equivalence proofs than \ToolARDiff{} across the different timeouts.
The primary reason for this difference is that \ToolARDiff{} only executes all paths in each of the two target programs once ($O(m+n)$).
\ToolPASDA{}, on the other hand, executes every path in the second program once for every path in the first program ($O(m\cdot{}n)$) when symbolically executing the product program described in Listing~\ref{lst:product-program}.
However, many paths are only partially explored because they are quickly identified to be unreachable, which significantly reduces the runtime impact caused by \ToolPASDA{}'s higher runtime complexity.
For \scNeq{} cases, \ToolARDiff{} takes around 2--3 times as long to provide non-equivalence proofs as \ToolPASDA{} does. This is mainly because \ToolARDiff{} only checks non-/equivalence after both programs have been fully explored by the symbolic execution. \ToolPASDA{}, on the other hand, checks non-/equivalence multiple times throughout its execution, i.e., every time a single program path has been fully explored. Consequently, \ToolPASDA{} can often provide a non-equivalence proof after having analyzed only a small subset of all program paths. %

Table~\ref{tab:runtime-results-per-step} shows the average runtimes of \ToolPASDA{}'s processing steps across the six analyzed timeout settings from 10~s to 3600~s. The largest factor contributing to \ToolPASDA{}'s overall runtime is the symbolic execution step, which takes an average of 4.5~s at the 10~s timeout setting and increases to 454.3~s at the 3600~s timeout setting. The partition classification and abstraction refinement steps also show noticeable runtime increases as timeouts are raised, reaching average runtimes of 115.4~s and 31.4~s, respectively, at the 3600~s timeout. Runtime requirements for the remaining steps of \ToolPASDA{}'s analysis process (i.e., initialization, instrumentation, program classification, and finalization) remain largely the same at around 1~s or less per step across the different timeout settings. As such, they only affect \ToolPASDA{}'s overall runtimes in relatively minor ways, particularly at higher timeouts where the runtime requirements of symbolic execution, partition classification, and abstraction refinement (all three of which involve the use of constraint solving) are much larger.

\begin{table}[tbp]
\captionsetup{justification=centering}
\caption{Average runtimes of \ToolPASDA{}s processing steps\\at the six evaluated timeout settings.} 
\label{tab:runtime-results-per-step}
\setlength{\tabcolsep}{4.5pt}
\begin{tabularx}{\columnwidth}{Xrrrrrr} \toprule
Step                     & \multicolumn{6}{c}{Ø Runtime (s)}                    \\ \cmidrule{2-7}
                         & \multicolumn{1}{c}{\rotatebox{90}{10~s}}   & \multicolumn{1}{c}{\rotatebox{90}{30~s}}    & \multicolumn{1}{c}{\rotatebox{90}{90~s}}    & \multicolumn{1}{c}{\rotatebox{90}{300~s}}   & \multicolumn{1}{c}{\rotatebox{90}{900~s}}    & \multicolumn{1}{c}{\rotatebox{90}{3600~s}}   \\ \midrule
Initialization           & 1.1 & 1.0  & 1.1  & 1.2  & 1.1   & 1.2   \\
Instrumentation          & 0.2 & 0.3  & 0.3  & 0.3  & 0.3   & 0.4   \\
Symb. Execution       & 4.5 & 11.4 & 26.4 & 67.1 & 147.3 & 454.3 \\
Part. Classification & 0.3 & 1.1  & 4.0  & 14.0 & 36.1  & 115.4 \\
Prog. Classification   & 0.0 & 0.0  & 0.0  & 0.0  & 0.0   & 0.0   \\
Abstr. Refinement               & 0.0 & 0.1  & 0.6  & 2.5  & 10.1  & 31.4  \\
Finalization             & 0.0 & 0.0  & 0.0  & 0.0  & 0.0   & 0.0   \\ 
\bottomrule
\end{tabularx}
\end{table}

\medbreak
\noindent\fbox{\begin{minipage}{0.97\columnwidth}
\textbf{Answer to RQ4}: 
For \scEq{} cases, \ToolARDiff{}, \ToolDSE{}, and \ToolSE{} generally have 10--50\% shorter runtimes than \ToolPASDA{}.
For \scNeq{} cases, \ToolSE{}'s runtimes are generally 5--15\% shorter than \ToolPASDA{}'s whereas the runtimes of \ToolARDiff{} and \ToolDSE{} are often multiple times as long. %
The processing step that makes up the largest portion of \ToolPASDA{}'s overall runtimes is the symbolic execution step, followed by the partition classification step and the abstraction refinement step. 
\end{minipage}}

\section{Discussion}
\label{sec:discussion}

In the following subsections, we discuss the benefits that \ToolPASDA{} provides compared to existing equivalence checking approaches as well as possible threats to the validity of our results and how we addressed them.

\subsection{Benefits of \ToolPASDA{}}
\label{sec:benefits-of-partition-level-data}

As shown by our evaluation, there are two main dimensions along which \ToolPASDA{} improves upon the current state-of-the-art for equivalence checking of software programs: (i)~\ToolPASDA{} provides non-/equivalence \emph{proofs} for a larger percentage of analyzed programs and partitions than existing tools (see RQ1 and RQ3), and (ii)~\ToolPASDA{} provides \emph{best effort} equivalence classifications of \scMaybeEq{} and \scMaybeNeq{} for some programs and partitions that existing tools simply classify as \scUnknown{} (see RQ2 and RQ3). \ToolPASDA{} achieves these improvements at the cost of runtimes that are moderately longer than those of existing tools (see RQ4). Increases in equivalence checking accuracy directly benefit use cases that depend on equivalence checking as part of their analysis process. Such use cases are quite diverse, including the verification of compiler optimizations \citep{dahiya2017black}, refactoring assurance for developers \citep{person2008differential}, test suite amplification \citep{danglot2019augmentation}, and many others beyond those (see, e.g., \cite{sun2016inteq,mora2018client,mercaldo2021equivalence}).

Regarding the potential benefits of best effort equivalence classifications, it is important to note that these classifications always arise as the result of a partial proof of non-/equivalence. After all, a program-level result of \scMaybeEq{} is reported when \ToolPASDA{} proves that the two target programs produce equivalent outputs for \emph{some} partitions but cannot provide a full equivalence proof because (i)~some outputs cannot be proven to be non-/equivalent (i.e., are \scUnknown{}) or (ii)~some partitions are not reached by \ToolPASDA{}'s analysis due to the used depth limit or timeout settings. Similarly, a result of \scMaybeNeq{} is reported when \ToolPASDA{} \emph{cannot} prove whether a specific path through the two target programs is reachable by a concrete execution of the product program but \emph{can} prove that some outputs are non-equivalent \emph{iff} the path \emph{is} reachable. As such, the introduction of best effort equivalence classifications leads to a more complete description of the overall program behavior than would be achieved without them.

Based on this understanding of best effort equivalence classifications as \emph{partial} non-/equivalence proofs, we envision the use cases for this information to be very similar to those for equivalence checking in general. For example, test suite amplification approaches \citep{danglot2019augmentation} could be adapted to generate tests not only for \scNeq{} partitions, but also for \scMaybeNeq{} partitions. If at least some of the \scMaybeNeq{} paths are reachable, this will ensure that more changed program behaviors are covered by the generated tests. If none are reachable, the generated number of tests would simply be higher without any tangible benefit, though the false positive \scMaybeNeq{}'s will be revealed as such upon execution of the tests. Similarly, spectrum-based fault localization formulas \citep{wong2016survey} could be adapted to use different weighting factors that consider the information provided by best effort classifications with a lower weight than the information provided by non-/equivalence proofs. While a detailed evaluation of such adaptations is beyond the scope of this paper, we plan to conduct a more thorough investigation of the potential benefits of best effort classifications for various existing use cases in future work.

\subsection{Threats to Validity}
\label{sec:threats-to-validity}

\paragraph{Construct Validity} Benchmarking is widely established as one of the main methodologies for comparing the runtime performance and accuracy of newly developed approaches to the state-of-the-art. We used a well established benchmark~\citep{badihi2020ardiff} that was also used by \ToolARDiff{} to measure the accuracy and runtime performance of our approach, thus demonstrating our approach's effectiveness and efficiency in an evidence-based and reproducible way.

\paragraph{Internal Validity} To mitigate threats to the internal validity of our results, we manually validated the outputs (program-level and partition-level equivalence classifications, partition-effects pairs, coverage information, etc.) produced by the four tools for a random sample of investigated cases. During this validation, we did not identify any outputs that we would deem incorrect. The results of this manual validation are further supported by the results of our evaluation, which demonstrate that neither of the analyzed tools produced false positive \scEq{} or \scNeq{} classifications for any of the analyzed cases. %

\paragraph{External Validity} Our findings might not generalize to cases beyond those that we investigated in our evaluation. To mitigate this threat, we conducted our evaluation based on the existing \ToolARDiff{} benchmark \citep{badihi2020ardiff}, which was itself constructed by combining and extending multiple smaller benchmarks for symbolic execution-based equivalence checking. Still, certain program constructs such as recursion, strings, and arrays, which are difficult for symbolic execution and automated decision procedures to reason about, are missing from the benchmark, which limits the generalizability of our results. In particular, our results are unlikely to generalize to (larger) real-world cases due to the inherent limitations of the underlying decision procedures (see Section~\ref{sec:symbolic-execution}), though we share this limitation with all existing equivalence checking approaches.

\section{Related Work}
\label{sec:related-work}

As described throughout this paper, we reuse several ideas that were originally proposed by the authors of \ToolDSE{} \citep{person2008differential}, \ToolARDiff{} \citep{badihi2020ardiff}, and \ToolPRV{} \citep{bohme2013partition} in our approach. For example, like \ToolDSE{}, we also replace unchanged parts of the source code with UIFs. Furthermore, like \ToolARDiff{}, we iteratively refine the set of introduced UIFs to improve overall equivalence checking accuracy. Finally, like \ToolPRV{}, we conduct a partition-based analysis of the two target programs. Additionally, \ToolPASDA{} provides best effort equivalence classifications for cases that it cannot prove to be either equivalent or non-equivalent, whereas existing approaches such as UC-KLEE~\citep{ramos2011practical}, PEQ\textsc{check}~\citep{jakobs2021peqcheck}, and many more that have been developed throughout the years (e.g., \cite{godlin2009regression, backes2013regression, beyer2013precision, felsing2014automating, fedyukovich2016property}) simply classify such cases as \scUnknown{}.

To further improve the utility of its outputs, \ToolPASDA{} provides supporting information alongside its (best effort) equivalence classification results. This information includes  partition-level equivalence checking results as well as execution traces and concrete and symbolic inputs and outputs. Similar information is provided by some existing equivalence checking approaches. For example, \ToolPRV{} \citep{bohme2013partition} also provides partition-level equivalence checking results. DSE \citep{person2008differential} and DiSE \citep{person2011directed, yang2014directed} both provide behavioral deltas for non-equivalent partitions. Furthermore, \citet{mercer2012computing} present an extension for Java Pathfinder \citep{havelund2000model, javapathfinder} that visualizes program statements that DiSE identifies to be impacted by identified changes via highlighting of source code and control flow graphs. Symdiff \citep{lahiri2012symdiff} provides a set of concrete inputs that demonstrates non-equivalence, and highlights the corresponding execution paths in the source code. \citet{partush2014abstract} use abstract interpretation \citep{cousot1977abstract} to characterize changed as well as unchanged program behavior.

Various other approaches exist that offer different representations of software changes but do not provide functional non-/equivalence proofs. For example, \citet{le2014patch} propose multiversion interprocedural control flow graphs to represent changes in control flow across any number of program versions. LSdiff offers mechanisms that can group related changes and describe identified inconsistencies in these groups \citep{kim2009discovering}. Dex applies graph differencing to semantic graphs of programs and calculates summary statistics from the differencing results \citep{raghavan2004dex}. Furthermore, software evolution management tools such as Diffbase \citep{wu2021diffbase} and EvoMe \citep{wu2021evome} provide platforms for processing, storage, and exchange of change information across different analyses, and offer search features to make access to this data more convenient and less time consuming~\citep{di2022diffsearch}. Integrating \ToolPASDA{}'s change information into such a tool might be a fruitful direction for future work. %

\section{Conclusions}
\label{sec:conclusions}

We presented \ToolPASDA{}, our partition-based semantic differencing approach with best effort classification of undecided cases. \ToolPASDA{} conducts multiple analysis iterations in which it aggregates partition-level equivalence checking results obtained by symbolically executing the product program created from two target programs to produce program-level equivalence classifications. If neither equivalence nor non-equivalence can be formally proven, \ToolPASDA{} employs a set of classification heuristics to produce a best effort equivalence classification instead. To further improve the utility of its results, \ToolPASDA{} aims to provide additional supporting information consisting of execution traces and concrete and symbolic inputs and outputs along with its equivalence classification results.

To evaluate \ToolPASDA{}'s equivalence checking accuracy and runtime performance, we used an existing equivalence checking benchmark \citep{badihi2020ardiff} and compared the results to three state-of-the-art equivalence checking approaches (i.e., \ToolDSE{} \citep{person2008differential}, \ToolARDiff{}, and \ToolPRV{} \citep{bohme2013partition}). \ToolPASDA{} was able to provide non-/equivalence proofs for 61--74\% of cases in the benchmark at timeout settings from 10~s to 3600~s, thus achieving equivalence checking accuracies that are 3--7\% higher than the best results achieved by the three existing tools. Furthermore, \ToolPASDA{}'s best effort classifications were correct for 70--75\% of equivalent and 55--85\% of non-equivalent cases across the different timeouts.

Our evaluation results demonstrate that \ToolPASDA{} can provide more complete descriptions of analyzed program pairs than existing approaches through (i) its increased equivalence checking accuracy and (ii) the best effort equivalence classifications that are newly introduced by it. Increases in equivalence checking accuracy, i.e., in the percentage of cases that can be proven to be non-/equivalent, directly benefit existing use cases that rely on equivalence checking such as test amplification \citep{danglot2019augmentation} and refactoring assurance for developers \citep{person2008differential}. Since best effort equivalence classifications always arise as the result of partial non-/equivalence proofs, we envision that they will benefit similar use cases, which we plan to investigate in more detail in future work.

\section{Data Availability}

A replication package that contains the implementations of the four evaluated tools (\ToolPASDA{}, \ToolARDiff{}, \ToolDSE{}, and \ToolPRV{}), the used benchmark cases, our evaluation scripts, and all evaluation results is available on Zenodo \citep{replicationpackage}. %

\bibliographystyle{elsarticle-harv} 
\bibliography{references}

\end{document}